\documentclass[11pt,a4paper,english,british]{article}
\usepackage{lmodern}

\usepackage[T1]{fontenc}
\usepackage[utf8]{inputenc}
\usepackage{float}
\usepackage{booktabs}
\usepackage{units}
\usepackage{amsmath}
\usepackage{amssymb}
\usepackage{esint}

\makeatletter

\pdfpageheight\paperheight
\pdfpagewidth\paperwidth

\providecommand{\tabularnewline}{\\}

\usepackage{jcappub}
\numberwithin{equation}{section}

\renewcommand\[{\begin{equation}}
\renewcommand\]{\end{equation}} 

\makeatother

\usepackage{babel}
\begin{document}

\title{Consistent perturbations in an imperfect fluid}

\author[a]{Ignacy Sawicki,}

\author[b]{Ippocratis D. Saltas,}

\author[a]{Luca Amendola}

\author[c]{and Martin Kunz}

\affiliation[a]{Institut für Theoretische Physik, Ruprecht-Karls-Universität Heidelberg\\
Philosophenweg 16, 69120 Heidelberg, Germany}

\affiliation[b]{Department of Physics and Astronomy, University of Sussex \\
Brighton, BN1 9QH, United Kingdom\\
and \\
School of Physics and Astronomy, University of Nottingham\\
Nottingham, NG7 2RD, United Kingdom }

\affiliation[c]{Départment de Physique Théorique and Center for Astroparticle Physics\\
Université de Genève\\
Quai E. Ansermet 24, CH-1211 Genève 4, Switzerland}

\emailAdd{ignacy.sawicki@uni-heidelberg.de, ippocratis.saltas@nottingham.ac.uk}

\emailAdd{l.amendola@thphys.uni-heidelberg.de, martin.kunz@unige.ch}

\abstract{We present a new prescription for analysing cosmological perturbations
in a more-general class of scalar-field dark-energy models where the
energy-momentum tensor has an imperfect-fluid form. This class includes
Brans-Dicke models, $f(R)$ gravity, theories with kinetic gravity
braiding and generalised galileons. We employ the intuitive language
of fluids, allowing us to explicitly maintain a dependence on physical
and potentially measurable properties. We demonstrate that hydrodynamics
is not always a valid description for describing cosmological perturbations
in \emph{general }scalar-field theories and present a consistent alternative
that nonetheless utilises the fluid language. \\
We apply this approach explicitly to a worked example: k-\emph{essence
}non-minimally coupled to gravity. This is the simplest case which
captures the essential new features of these imperfect-fluid models.
We demonstrate the generic existence of a new scale separating regimes
where the fluid is perfect and imperfect. We obtain the equations
for the evolution of dark-energy density perturbations in both these
regimes. The model also features two other known scales: the Compton
scale related to the breaking of shift symmetry and the Jeans scale
which we show is determined by the speed of propagation of small scalar-field
perturbations, i.e. causality, as opposed to the frequently used definition
of the ratio of the pressure and energy-density perturbations.}

\maketitle

\section{Introduction\label{sec:Introd}}

Many models, better or worse motivated by fundamental physics, have
been proposed as an alternative to the cosmological constant for providing
a mechanism to accelerate the expansion of the universe at late times.
Perhaps the simplest, and therefore most studied, are models which
feature a single extra scalar degree of freedom, which is treated
as classical. The archetypal example of this dark energy (``DE'')
are quintessence models \cite{Ratra:1987rm,Wetterich:1987fm}, where
a canonical scalar field rolls in a potential resulting a time-varying
equation of state and allowing for (small) inhomogeneities in the
dark-energy fluid. On the other hand, $f(R)$ gravity models \cite{Carroll:2003wy},
where the Einstein-Hilbert action for gravity is modified, allow for
a different response of the metric to the presence of matter from
that in general relativity (``GR'') and usually are discussed as
examples of modified gravity. However, these models can also be reformulated
as a particular subclass of Brans-Dicke theories featuring a single
extra scalar, albeit non-minimally coupled to gravity \cite{Brans:1961sx,Chiba:2003ir}.
It was quickly understood that whereas the background expansion history
can be indistinguishable between various classes of these models,
the evolution of dark-matter perturbations can differ significantly
and therefore formation of large-scale structure provides the key
to understanding the role played by these theories, if any, in accelerating
the universe \cite{Ishak:2005zs}.

k-\emph{essence} was introduced as an extension of quintessence models
to non-canonical kinetic terms \cite{ArmendarizPicon:1999rj,ArmendarizPicon:2000ah,ArmendarizPicon:2000dh}
and can be interpreted as a perfect fluid with a speed of sound different
from that of light \cite{Garriga:1999vw}. This allows for a Jeans
scale smaller than the cosmological horizon and therefore for DE which
can cluster significantly. In particular, the discovery of generic
attractors where the equation of state approaches that of the vacuum
while the sound speed tends to zero \cite{ArkaniHamed:2003uy} also
spurred the discussion of such models as describing both dark energy
and dark matter through a single degree of freedom \cite{Scherrer:2004au,Bertacca:2007ux,Lim:2010yk}.
One tends to think of these dark-energy models as perfect fluids,
obeying hydrodynamics and studies the evolution of linear perturbations
by considering the conservation of the perturbed energy-momentum tensor
\cite{Bardeen:1980kt,Ma:1995ey,Hu:1998tj}.

On the other hand, when studying large-scale structure in $f(R)$
theories \cite{Song:2006ej,Tsujikawa:2007xu}, one tends to think
in terms of a modification of the response of the two scalar gravitational
potentials, $\Phi$ and $\Psi$, to matter perturbations. In particular
the generic result for this class of theories is that the lensing
potential, $\Phi-\Psi$, is ``unaffected'' and depends only on the
matter perturbations and an evolving Newton's constant, while the
modification of gravity creates anisotropic stress giving $2\Phi+\Psi=0$.%
\footnote{For an investigation of the relation between anisotropic stress and
cosmological evolution in non-linear gravity models see ref \cite{Saltas:2010tt}.%
} These relationships are simple and therefore are frequently generalised
in order to constrain such theories from data. A large number of phenomenological
parameterisations of models of modified gravity exist \cite{Linder:2005in,Knox:2005rg,Ishak:2005zs,Kunz:2006ca,Laszlo:2007td,Caldwell:2007cw,Amendola:2007rr,Pogosian:2010tj}
and usually focus on directly modifying the relationships between
the potentials and matter perturbations without necessarily asking
what this implies for the underlying model. Considering the lack of
well-motivated models, this might not be considered a disadvantage.
A different approach, based on symmetries of the Einstein equations
without a prior reference to a particular dark-energy model, was employed
in ref.~\cite{Battye:2012eu}, while a parameterised post-Friedmann
approach was proposed in \cite{Hu:2007pj} and further developed in
\cite{Baker:2011jy}. 

$f(R)$ models and this sort of modifications of gravity are quite
constrained by Solar-System tests as a result of their being equivalent
to fifth forces. One can find viable models \cite{Hu:2007nk,Starobinsky:2007hu,Amendola:2006we}
which can evade the constraints through the chameleon effect \cite{Khoury:2003aq,Khoury:2003rn},
or one can propose that the extra scalar does not couple to baryons,
but only to dark matter \cite{Wetterich:1994bg,Amendola:1999er}.
These types of interacting DE models can have dynamics on large scales
very similar to modified gravity models, since the contribution of
baryons to the energy density is very subdominant. However, philosophically
they are no longer ``modified gravity'' since the coupling is not
universal --- it no longer couples purely to the energy-momentum tensor.
On the other hand, the full breadth of such models is much wider than
for $f(R)$ theories, which are but a subclass. When studying interacting
dark energy, one tends to treat the scalar field differently from
matter: the equation of motion for the scalar is usually parameterised
and solved, while the dark matter and baryons are treated as a hydrodynamical
fluid. 

Another class of scalar-field models was discovered by considering
gravity modifications arising from a potential higher-dimensional
embedding of our universe \cite{Dvali:2000hr,Deffayet:2000uy,Gabadadze:2009ja}.
In the decoupling limit \cite{Luty:2003vm,Nicolis:2004qq,Gabadadze:2006tf},
this DGP gravity reduces to a scalar theory, where the scalar Lagrangian
contains second derivatives of the scalar field and therefore is not
in the k-\emph{essence} class. A further generalisation of such scalar-field
theories featuring second derivatives was then obtained in refs \cite{Nicolis:2008in,Nicolis:2009qm}
and named the \emph{galileon}. Such theories are not consistent on
general curved backgrounds, and the appropriate covariantisation was
obtained in \cite{Deffayet:2009wt,Deffayet:2009mn}. Kinetic gravity
braiding, a generalisation of the DGP decoupling limit, was independently
found in refs \cite{Deffayet:2010qz,Kobayashi:2010cm} and soon it
was proven that the most general theory of a single scalar with no
more than second derivatives is a similar generalisation of the galileons
\cite{Deffayet:2011gz}. Indeed these generalised galileons were discovered
a long time ago \cite{Horndeski:1974} and largely ignored. The higher-dimensional
origin of the galileon theories was studied in refs \cite{deRham:2010eu,Hinterbichler:2010xn,Goon:2011qf,Goon:2011uw},
while the galileon theories also have appeared in the decoupling limit
of consistent massive-gravity as actions for the helicity-0 mode of
the massive graviton \cite{deRham:2010ik,deRham:2010kj,Hassan:2011vm}.

These theories all contain second derivatives in the action and therefore
have energy-momentum tensors (``EMT'') which do \emph{not }have
the form of a perfect fluid \cite{Pujolas:2011he,Gao:2011mz} and
therefore provide for new phenomenology. For example, the null-energy
condition (``NEC'') can be violated stably, resulting in explicitly
ghost-free phantom DE models or even permitting the construction of
initial phases of evolution of the universe alternative to inflation
\cite{Creminelli:2010ba,Qiu:2011cy,Easson:2011zy}. Equations for
linear cosmological perturbations for the most general of such theories
were derived in \cite{Gao:2011qe,DeFelice:2011hq}.\textbf{ }Indeed,
models with stabilised ``phantom-crossing'' and subsequent evolution
were discussed from the point of view of the effective action for
perturbations already in \cite{Creminelli:2006xe,Creminelli:2008wc}:
adding higher-derivative operators acts to stabilise configurations
violating the NEC.\\

What strikes in the above discussion is the multitude of approaches
that are used to study perturbations, each chosen depending on the
underlying Lagrangian for the scalar. However, all the models discussed
here are structurally extremely similar. A common framework for describing
perturbations would allow for a single consistent method of constraining
the class of models of DE comprising a single scalar. One could argue
that solving the equation of motion for the scalar provides exactly
such a framework. However, if we assume that the DE cannot be studied
directly, but only through its effect on the gravitational field in
which baryons and light propagate, the equation of motion contains
too much information. Gravity couples only to the EMT and therefore
will mostly depend on the fluid properties of the DE and not on the
precise value of the scalar field. 

The case of k-\emph{essence} is a pertinent example. Any particular
choice within this class of Lagrangians gives a particular expansion
history for the cosmological background, with a particular evolution
of the equation-of-state parameter, $w$. This solution (based on
the choice of Lagrangian) then also determines the speed of propagation
of the scalar-field perturbations, $c_{\text{s}}^{2}$. Given the
history of these two variables, the evolution of linear energy-density
perturbations of a k\emph{-essence }DE\emph{ }is \emph{completely}
determined at all scales. Any k-\emph{essence} model which happens
to have the same history of evolution of $w$ and $c_{\text{s}}^{2}$
will have exactly the same evolution of linear perturbations, even
if the two Lagrangians are very different and only happen to have
coincided in this way on two individual solutions. There is therefore
no way to reconstruct a Lagrangian even limited to this single class
of theories, even if we ignore the fact that the data that can be
obtained will only ever be limited to some maximal precision and we
will only be able to measure the properties of the DE indirectly.
Unless DE interacts directly with baryonic matter in a way that is
not equivalent to redefining a gravitational metric, we will only
be able to measure the properties of the DE perturbations indirectly
by looking at scale-dependent modifications of the DM power spectrum.
Given this, the best we can hope for is to measure $w$ through its
effect on geometry and the sound speed of DE by detecting some sort
of scale of transition in behaviour that would be consistent with
a Jeans length \cite{Sapone:2009mb,Sapone:2010uy}.

A similar situation occurs when dark energy is described by the class
of $f(R)$ theories. Here, there appears to be only one scale, that
of the (time-varying) Compton wavelength of the scalar, the chameleon
effect allowing for the restoration of GR in dense regions through
non-linearities notwithstanding. On both sides of this scale, the
evolution of linear perturbations is completely determined by the
fact that we have chosen to consider this limited class of Lagrangians,
just as it was for k-\emph{essence}. Indeed, specifying the evolution
of $w$ and of the Compton scale gives us the totality of the information
that can be gleaned from the background and the linear perturbations.

In fact, the standard discussion of $f(R)$ gravity neglects two more
scales which are implicitly contained in the theory. Firstly, there
is still the Jeans scale. However, in this case, the speed of sound
is equal to that of light and therefore the Jeans scale lies at the
cosmological horizon and is thus unobservable. Secondly, there exists
a new scale controlling whether the fluid is perfect of imperfect,
which will be the subject of much of this paper. $f(R)$ models are
a limit where this scale lies at infinity and therefore it is also
unobservable there. \\

The main aim of this paper is to provide a prescription for describing
linear perturbations which as far as possible depends only on physical
properties of the model, such as the equation of state or the sound
speed, which one could hope to measure. Our approach is to describe
the DE through the conservation equations for its EMT, rather than
the equation of motion. This provides the evolution equations for
the energy-density perturbations which are then directly connected
to the gravitational potentials. For any theory, these equations can
be solved provided we supply two closure relations: relationships
of the pressure perturbation and anisotropic stress to the energy-density
perturbation. The solutions for linear perturbations of any fluid
are always fully determined by these two functions.%
\footnote{Some EMT non-conservation issues related to the evolution of the effective
Planck mass notwithstanding.%
} The question we answer is how to properly obtain these relationships
for a model featuring a single scalar degree of freedom.

We show explicitly that even for general k-\emph{essence} models these
closure relations, which are well known \cite{Kunz:2006wc,Christopherson:2008ry},
are not hydrodynamical and one should never a priori assume hydrodynamical
behaviour in more general cases. However, on any time slice all components
of the EMT are determined by the values of the degrees of freedom
at that moment in time. Thus, for any class of models based on degrees
of freedom with classical equations of motion, we can obtain such
closure relations. Indeed these relations define the totality of the
properties of the dark energy at the linear level of perturbations.
We provide a form of these closure relations for any scalar-field
theory and we explicitly calculate them for k-\emph{essence} coupled
non-minimally to gravity. This particular class contains both k-\emph{essence
}and $f(R)$ theories and as a result of the non-minimal coupling
to gravity features second-derivatives of the scalar in the EMT in
the Jordan frame. In this sense, it mimics the form of the EMT possessed
by more-general galileon theories and therefore is a useful example.

We show that in general scalar-field models we have three scales:
the Jeans scale, the Compton scale and a new scale determining whether
the fluid is in a perfect or imperfect regime. This is \emph{in addition
}to the cosmological-horizon scale. These three scales are essentially
independent and can lie in any order and are determined by the Lagrangian
class and the particular background solution. This new transition
scale appears in all scalar-field models more general than k-\emph{essence}
and is determined by the relative importance of the second-derivative
terms and the k-\emph{essence} terms in the perturbed EMT. For example,
it is only inside this scale that the DE fluid can carry anisotropic
stress at all. This scale is not visible in the equation of motion
for the scalar, but only appears when the EMT is considered.

As an example of the power of our prescription, we obtain the equations
for the evolution of energy-density perturbations for the non-minimally
coupled \emph{k-essence }DE model and solve them analytically under
the assumption that the background evolution exhibits scaling behaviour
(i.e. has a constant equation-of-state parameter).\\

The paper is structured as follows: we start in section \ref{sec:FluidFormalism}
by describing a formalism for decomposing the EMT into the fluid variables
covariantly. This description allows us to obtain exact forms for,
among others, the energy density and pressure in terms of the scalar-field,
which can then be perturbed directly. In this way, we obtain the conservation
equations for the linear perturbations of the EMT in section \ref{sub:EvolEqLin}.
In section \ref{sub:hydro} we discuss what hydrodynamics implies
for the relationship between pressure and energy density in perfect
fluids, and by deriving this relationship for general k-\emph{essence
}models in section \ref{sub:k-ess} we show that the hydrodynamical
relations are not obeyed. We then propose a form for these closure
relations which would be obeyed by any DE model comprising a single
scalar field in section \ref{sub:GenPerts}. In section \ref{sec:BDK},
we turn to a non-minimally coupled k\emph{-essence }as worked\emph{
}example. We discuss the general properties of this theory, before
turning to calculate the closure relations between the pressure, energy
density and anisotropy perturbations in section \ref{sub:DEonly_perts}.
We demonstrate explicitly how to calculate them and show that they
are of the form contained within the general parameterisation introduced
earlier. We explicitly demonstrate the existence of the three scales
in the problem and derive the equation for the evolution of the density
contrast in these models. We summarise our findings in section \ref{sec:Conc}.

\section{Fluid formalism\label{sec:FluidFormalism}}

It is well known that theories such as k-\emph{essence }can be much
more intuitively described in terms of relativistic hydrodynamics
\cite{ArmendarizPicon:1999rj}. We are going to employ this language
extensively and show that it provides a very natural framework in
which to discuss theories with contributions to the energy-momentum
tensor containing second derivatives. This language was also previously
employed to study scalar field theories with kinetic gravity braiding
\cite{Pujolas:2011he} (to where we refer the reader for a more detailed
exposition of this formalism), but was developed for relativistic
potential flows much earlier \cite{Landafshitz_Fluid,Moncrief:1980,DiezTejedor:2005fz,Bilic:2008zk}.
We will proceed in this manner, leaving aside for the moment the question
of whether the more general scalar-field theories we would like to
describe are in fact hydrodynamical fluids.

\subsection{Standard definitions}

In this discussion, we are going to be describing a scalar-field theory.
Following standard practice, we will define a canonical kinetic term
for the scalarIn cosmology one usually takes gradients as being time-like.
It is only with this choice that the fluid description can be valid,
thus we will always assume that $X>0$. The scalar-field gradient
now gives a time-like vector field, which when appropriately normalised
can be identified with a velocity field,
\begin{equation}
u_{\mu}\equiv-\frac{\partial_{\mu}\phi}{\sqrt{2X}}\,.\label{eq:u_def}
\end{equation}
This velocity field is usually referred to as the fluid rest-frame,
but in the case of the more general models we discuss, this is not
always the case: there remains a non-vanishing energy flow in the
frame moving with the velocity $u^{\mu}$. We will therefore call
the frame defined by $ $(\ref{eq:u_def}) the \emph{scalar frame.
}We comment further on this frame choice in appendix \ref{sub:boosts}.

Given the velocity field, we can define a derivative along $u^{\mu}$
(a material derivative)
\begin{equation}
u^{\mu}\nabla_{\mu}=\frac{\mathrm{d}}{\mathrm{d}\tau}\,,\label{eq:time_deriv_def}
\end{equation}
thus making $\tau$ the proper time of an observer comoving with the
scalar frame. We will assume that this time derivative be positive,
and therefore $u^{\mu}$ be future directed. This allows us to think
of $\phi$ as a clock.

We can then proceed and define the transverse projector,
\begin{equation}
\perp_{\mu\nu}=g_{\mu\nu}+u_{\mu}u_{\nu}\,,\label{eq:perp_def}
\end{equation}
which plays the role of the first fundamental form in the hypersurfaces
$\Sigma_{\phi}:\phi(x)=\mathrm{const}$, i.e. the spatial metric for
the scalar-frame observer. The projector (\ref{eq:perp_def}) allows
us to decompose vectors and gradients into time and spatial parts
as observed in the scalar frame,
\begin{equation}
\nabla_{\mu}=-u_{\mu}u^{\lambda}\nabla_{\lambda}+\perp_{\mu}^{\lambda}\nabla_{\lambda}=-u_{\mu}\frac{\mathrm{d}}{\mathrm{d}\tau}+\overline{\boldsymbol{\nabla}}_{\mu}\,.\label{eq:deriv_decomp_def}
\end{equation}
We will refer to\textbf{ $\overline{\boldsymbol{\nabla}}_{\mu}$ }as
the spatial derivative. We can then proceed to decompose the gradient
of the vector field $u^{\mu}$, by defining the acceleration, a spatial
vector field
\begin{equation}
a_{\mu}\equiv\frac{\mathrm{d}}{\mathrm{d}\tau}u_{\mu}=u^{\lambda}\nabla_{\lambda}u_{\mu}\,,\label{eq:accel_def}
\end{equation}
which can also be expressed in terms of the scalar field
\begin{equation}
a_{\mu}=-\perp_{\mu}^{\lambda}\nabla_{\lambda}\ln\frac{\mathrm{d\phi}}{\mathrm{d}\tau}\,,\label{eq:a_as_m}
\end{equation}
and performing the standard kinematical decomposition
\begin{equation}
\nabla_{\mu}u_{\nu}=-u_{\mu}a_{\nu}+\sigma_{\mu\nu}+\frac{1}{3}\perp_{\mu\nu}\theta\,,\label{eq:grad_u}
\end{equation}
 where $\theta\equiv\nabla_{\mu}u^{\mu}$ is the expansion and 
\begin{equation}
\sigma_{\mu\nu}\equiv\perp_{(\mu}^{\lambda}\nabla_{\lambda}u_{\nu)}-\frac{1}{3}\perp_{\mu\nu}\theta\label{eq:shear_def}
\end{equation}
is the fully spatial (transverse) and traceless symmetric shear tensor.
The rotation tensor (the antisymmetric spatial part of (\ref{eq:grad_u}))
vanishes by virtue of the Frobenius theorem, since $u_{\mu}$ is a
gradient of a scalar.

In ref.~\cite{Pujolas:2011he}, it was proposed that instead of the
canonical scalar kinetic term, $X$, a more physical variable is in
fact
\begin{equation}
m\equiv\frac{\mathrm{d}\phi}{\mathrm{d}\tau}=u^{\mu}\nabla_{\mu}\phi\,.\label{eq:def_m}
\end{equation}
On the cosmological background, which will be the focus of our discussion,
the variable $m$ reduces to the derivative of the scalar with respect
to the coordinate time, $\mathrm{d}\phi/\mathrm{d}t$. It will also
be helpful to give the decomposition of the second derivative in terms
of the kinematical variables:
\begin{equation}
\nabla_{\mu}\nabla_{\nu}\phi=-\frac{\mathrm{d}m}{\mathrm{d}\tau}u_{\mu}u_{\nu}-m\left(a_{\mu}u_{\nu}+u_{\mu}a_{\nu}-\sigma_{\mu\nu}-\frac{1}{3}\perp_{\mu\nu}\theta\right)\,.\label{eq:Bmunu}
\end{equation}
In the hydrodynamical picture, the variable $m$ can be interpreted
as playing the role of the chemical potential for the charges which
make up the fluid. In this picture, a scalar-field theory is a thermodynamical
system at zero temperature, with $m$ acting as a chemical potential
and any dependence on $\phi$ representing an explicit dependence
of the properties on the internal clock. As argued in \cite{Akhoury:2008nn,Arroja:2010wy,Unnikrishnan:2010ag},
scalar-field theories such as k\emph{-essence }are perfect hydrodynamical
fluids \emph{only }when they are shift symmetric, with the action
invariant under the transformation $\phi(x)\rightarrow\phi(x)+\text{const}$,
i.e. hydrodynamics is only valid when there is no explicit dependence
on the clock carried by the fluid. We will return to this point in
the discussion of the validity of hydrodynamics in sections \ref{sub:hydro}
and \ref{sub:k-ess}. 

The configuration of a minimally coupled k-\emph{essence }theory is
fully specified by the pair $(\phi,m)$. Theories with second derivatives
in the Lagrangian, such as kinetic gravity braiding \cite{Pujolas:2011he}
are somewhat more complicated, since their configuration can also
depend on quantities such as the expansion $\theta$. Typically, the
energy momentum tensor of these theories will contain second derivatives
of the scalar. We will discuss this sort of example in section \ref{sec:BDK}.

\subsection{Evolution equations from EMT conservation\label{sub:EvolEqLin}}

At this stage, it is useful to jump ahead and be a little more specific.
In scalar models coupled non-minimally to gravity, the fundamental
Planck mass, $M_{\text{Pl}}$ will in general be modified by a function
of the scalar, let us call it $\varkappa$, through a coupling $e^{\varkappa}R$
in the Lagrangian. This would mean that the energy momentum tensor
(``EMT'') for the DE would in general contain a term proportional
to the Einstein tensor, $(1-e^{\varkappa})G_{\mu\nu}$. When the Einstein
tensor is solved for through the Einstein equations, the energy-momentum
tensor for the dark energy depends explicitly on the dark-matter configuration,
which we would like to think of as a separate degree of freedom with
its own dynamics (in fact, potentially multiple degrees of freedom:
dark matter, baryons, radiation, etc.). The prescription we introduce
in this paper calls for separating out this dependence, so as to leave
the DE EMT as much as possible independent of the particular configuration
of the external matter.\textbf{ }We will therefore explicitly solve
for the Einstein tensor and introduce an EMT, $T_{\mu\nu}^{X}$ with
the Einstein tensor removed. The Einstein equations are then given
by 
\begin{equation}
G_{\mu\nu}=T_{\mu\nu}^{X}+e^{-\varkappa}T_{\mu\nu}^{\text{ext}}\,,\label{eq:Einstein!+f}
\end{equation}
where we have chosen units where $M_{\text{Pl }}^{2}=8\pi G_{\text{N}}=1$.
$T_{\mu\nu}^{\text{ext}}$ is the possible EMT external to the dark-energy
system, for example representing dark matter, which would be explicitly
conserved, $\nabla^{\mu}T_{\mu\nu}^{\text{ext}}=0$. All the fluid
variables relating to the external matter will appear suppressed by
$e^{-\varkappa}$. The price for such a definition is the fact that
$T_{\mu\nu}^{X}$ is not conserved in the presence of matter external
to the system,
\begin{equation}
\nabla^{\mu}T_{\mu\nu}^{X}=e^{-\varkappa}T_{\mu\nu}^{\text{ext}}\nabla^{\mu}\varkappa\,.\label{eq:EMTCons}
\end{equation}
In this paper, will consider the isolated dark-energy system, setting
$T_{\mu\nu}^{\text{ext}}=0$, only making comments in passing when
the presence of external matter modifies significantly our results.
We will return to study the combined dark-energy-dark-matter system
in another work.

A natural question to ask is why we are not performing the analysis
in the Einstein frame, which is the typical way in which non-minimally
coupled theories are usually discussed \cite{Bean:2006up}. The Einstein
frame defined as the frame where the kinetic terms of the metric and
of the scalar field are unmixed. Indeed \emph{for the model }we present
in the worked example in section \ref{sec:BDK}, a conformal transformation
can be performed and the analysis of such a setup is standard: in
the Einstein frame we are dealing with a normal perfect fluid, albeit
non-minimally coupled to external matter. However, for more general
models of DE (kinetic gravity braiding, generalised galileons) no
general redefinition of frame where the scalar EMT always has perfect-fluid
form can be found. Such a diagonalisation can sometimes only be performed
only on the level of linear perturbations while the full non-linear
theory must always be written in a frame that mixes the two degrees
of freedom. Our motivation is that the simple model presented in our
worked example, in the Jordan frame, has many features in common with
the these general galileon models and therefore our prescription as
employed in the Jordan frame will transfer directly to more complex
situations. 

In general, the baryons, DM and gravity could all have different frames
in which they have minimal couplings. We can always pick one of those
frames as the basis for the analysis. Since most of our observations
are to do with baryons, it simplifies matters to assume that they
move on geodesics of some metric. This is the metric in which they
are minimally coupled. But in this frame, the DM and gravity could
both have some sort potentially different non-minimal coupling for
the scalar.\\

Once we have defined a velocity field, we can decompose the EMT into
the fluid variables that will be seen by the observer $\mathcal{O}$
comoving with velocity $u^{\mu}$:
\begin{equation}
T_{\mu\nu}^{X}=\mathcal{E}u_{\mu}u_{\nu}+\perp_{\mu\nu}\mathcal{P}+u_{\mu}q_{\nu}+u_{\nu}q_{\mu}+\tau_{\mu\nu}\,,\label{eq:Tmunu_decomp}
\end{equation}
where $\mathcal{E}\equiv T_{\mu\nu}^{X}u^{\mu}u^{\nu}$ is the energy
density observed by $\mathcal{O}$, $\mathcal{P}\equiv\perp^{\mu\nu}T_{\mu\nu}^{X}/3$
is the pressure, $q_{\lambda}\equiv-T_{\mu\nu}^{X}\perp_{\lambda}^{\mu}u^{\nu}$
is the energy flow as seen by $\mathcal{O}$ and $\tau_{\lambda\rho}\equiv T_{\mu\nu}^{X}\left(\perp_{\lambda}^{\mu}\perp_{\rho}^{\nu}-\frac{1}{3}\perp_{\lambda\rho}\perp^{\mu\nu}\right)$
is the spatial and traceless viscous-stress tensor.%
\footnote{The quantity $q_{\lambda}$ is usually called the heat flow. However,
as was demonstrated in \cite{Pujolas:2011he} for kinetic gravity
braiding theories, entropy production and heat flow are zero while
$q_{\lambda}$ does not vanish. In that case, the vector $q_{\lambda}$
quantifies the energy carried by charges moving in the scalar frame.%
} The observed values of the fluid quantities are not independent of
the choice of observer and are not just gauge effects. We discuss
this in Appendix \ref{sub:boosts}. As implied by the decomposition
of the second derivative of a scalar eq. (\ref{eq:Bmunu}), EMTs containing
second derivatives will in general have contributions to all the components
of the EMT, implying that the EMT has imperfect-fluid form: it is
impossible to find a velocity frame in which it could be decomposed
and possess a perfect-fluid form.

At least for some models (in particular, the one we discuss as a worked
example in section \ref{sec:BDK}), we can rewrite the energy flow
$q_{\lambda}$ and viscous stress $\tau_{\mu\nu}$ in terms of a scalar
potential with an appropriate number of derivatives:
\begin{eqnarray}
q_{\lambda} & = & \perp_{\lambda}^{\mu}\nabla_{\mu}q,\label{eq:potentials}\\
\tau_{\lambda\rho} & = & \left(\perp_{\lambda}^{\mu}\perp_{\rho}^{\nu}-\frac{1}{3}\perp_{\lambda\rho}\perp^{\mu\nu}\right)\nabla_{\mu}\nabla_{\nu}\pi\nonumber 
\end{eqnarray}
with $q$ the potential for energy flow and $\pi$ the anisotropic
stress.

We present a detailed discussion of the conservation equations for
the EMT using the language of covariant decomposition in Appendix
\ref{sec:CovDecomp}. Here we are just going to discuss the linear
scalar perturbations on a spatially flat Friedmann-Lemaître-Robertson-Walker
(``FLRW'') metric working in the Newtonian gauge, assuming that
vector and tensor perturbations be negligible,
\begin{equation}
\mathrm{d}s^{2}=-(1+2\Psi)\mathrm{d}t^{2}+a^{2}(t)(1+2\Phi)\mathrm{d}\boldsymbol{x}^{2}\,.\label{eq:metric}
\end{equation}
The exact equations presented in the appendix, (\ref{eq:Tcons_u_scalar})
and (\ref{eq:Tcons_perp_scalar}), can be perturbed directly. Doing
so, we obtain the standard results. At the background level only the
timelike equation contributes, giving the energy-conservation equation,
\[
\dot{\mathcal{E}}+3H(\mathcal{E}+\mathcal{P})=0\,,
\]
with the overdot representing the derivative with respect to coordinate
time $t$. At the linear level in perturbations, we can transform
into momentum space obtaining:
\begin{flalign}
 & \dot{\delta\mathcal{E}}+3H\left(\delta\mathcal{E}+\delta\mathcal{P}\right)+\left(\mathcal{E}+\mathcal{P}\right)\left(\Theta+\dot{\Phi}\right)-\frac{k^{2}\delta q}{a^{2}}+\dot{q}\Theta=0\,,\label{eq:EnCons_sc}\\
 & \left(\mathcal{E}+\mathcal{P}\right)\left(\dot{\Theta}+2H\Theta-\frac{k^{2}}{a^{2}}\Psi\right)-\frac{k^{2}}{a^{2}}\delta\mathcal{P}+\Theta\dot{\mathcal{P}}+\label{eq:MOmCons_sc}\\
 & -3H\left(\frac{k^{2}}{a^{2}}\delta q-\dot{q}\Theta\right)-\frac{1}{a^{2}}\left(k^{2}\delta q-a^{2}\dot{q}\Theta\right)^{\cdot}-\frac{2}{3}\frac{k^{4}}{a^{4}}\delta\pi=0\,,\nonumber 
\end{flalign}
where the $\delta$ prefix signifies a perturbed quantity and we have
defined the divergence of the perturbation of spatial velocity for
the observer in the scalar frame:%
\footnote{Note that there is a relative factor of $a$ between $\Theta$ defined
above and the variable $\theta$ used by Ma and Bertschinger \cite[Eq. (23b)]{Ma:1995ey},
so that $\theta_{\text{MB}}=a\Theta$. This is a result of our using
coordinate instead of conformal time.%
} 
\begin{equation}
\Theta\equiv ik_{i}\delta u^{i}=\frac{k^{2}\delta\phi}{ma^{2}}\,.\label{eq:ThetaX_def}
\end{equation}
It is customary to make the equations (\ref{eq:EnCons_sc}) and (\ref{eq:MOmCons_sc})
dimensionless by dividing through by the background energy density:
for us, it will prove more convenient to keep them in this form, since
we do not wish to at this stage follow the usual practice of defining
a sound speed, etc. and we wish to avoid any potential singularities
this procedure might introduce in the variables.\\

In the Newtonian gauge, the metric potentials $\Phi$ and $\Psi$
are coincident with the gauge invariant Bardeen potentials \cite{Bardeen:1980kt}.
This means that if one replaced all the perturbed variables with their
Bardeen gauge-invariant correspondents, the form of the equations
(\ref{eq:EnCons_sc}) and (\ref{eq:MOmCons_sc}) would not change.\textbf{
}Gauge invariance, however, does not imply frame invariance. The standard
references (e.g. ref.~\cite{Ma:1995ey}) perform the decomposition
of the EMT eq.~(\ref{eq:Tmunu_decomp}) in the rest frame of the
fluid and therefore do not contain the energy flow terms involving
the potential $q$. The difference between any two frames is a Lorentz
boost between the two velocities of the observers in whose rest frame
the EMT is decomposed. Given that the background is FLRW, there is
only one reasonable frame on the background level, that of the background
comoving observer. Thus all the frames differ purely on the level
of perturbations, with the boosts transforming equations (\ref{eq:EnCons_sc})
and (\ref{eq:MOmCons_sc}) by trading the divergence of energy flow
$k^{2}\delta q$ for a redefinition of the divergence of the momentum
perturbation $(\mathcal{E}+\mathcal{P})\Theta$.

For a model with second derivatives in the EMT, the choice of any
of these related frames is perfectly fine \emph{apart} from the rest
frame. In vacuum configurations, $\mathcal{E}+\mathcal{P}=0$ and
the divergence of momentum perturbation $(\mathcal{E}+\mathcal{P})\Theta$
vanishes. Therefore the rapidity of the boost required to compensate
a divergence of energy flow $k^{2}\delta q$ tends to infinity as
we approach the vacuum equation of state. This means that close to
vacuum configurations, perturbation theory in terms of the underlying
scalar-field variables itself breaks down if we require the frame
to be a rest frame.\textbf{ }However, we should emphasise that this
is merely a problem of the particular choice of variables and not
of perturbation theory in general. 

We have provided more detail for this discussion in Appendix \ref{sub:boosts}.
Suffice it to say that the presence of the energy flow potential $q$
in the perturbation equations (\ref{eq:EnCons_sc}) and (\ref{eq:MOmCons_sc})
will be the origin of much of the novel behaviour described in this
paper. Scalar-field theories more general than k-\emph{essence} always
contain such a term in the EMT decomposed in the scalar frame (\ref{eq:u_def}).
Close enough to vacuum configurations, the coefficients of the $\Theta$
terms in (\ref{eq:EnCons_sc}) and (\ref{eq:MOmCons_sc}) driving
the evolution of perturbations under normal circumstances vanish and
that role is taken over by the $q$ terms. This then allows for such
behaviour as a stable violation of the null-energy condition \cite{Creminelli:2006xe,Creminelli:2008wc,Deffayet:2010qz,Easson:2011zy}
or ``phantom crossing''. The existence of the additional derivatives
on the $\delta q$ perturbations compared to the $\Theta$ terms will
mean that the transition between the two behaviours will not only
depend on the equation of state of the dark energy but also on scale. 

Another aspect worth noting here is that the background value of $q$
appears in eqs (\ref{eq:EnCons_sc}) and (\ref{eq:MOmCons_sc}). This
does not have to vanish, only its spatial gradient must, at the background
level (the total value of $q$ is some function of the scalar field,
$\phi$ and $m$, so its background value is going to be that function
of the background values of $\phi$ and $m$). \\

The aim of this discussion is to reduce the perturbation equations
(\ref{eq:EnCons_sc}) and (\ref{eq:MOmCons_sc}) into a single evolution
equation for the density perturbation $\delta\mathcal{E}$. In the
rest frame, this is simple since one can just eliminate the variable
$\Theta$. In the scalar frame, in general both $\Theta$ and $\delta q$
have to be eliminated. Luckily, one can exploit the structure of the
conservation equations arising from their transformations under Lorentz
boosts of $u^{\mu}$ to make an \emph{algebraic }redefinition of $\Theta$
to define the divergence of the total energy flow, 
\begin{equation}
\Xi\equiv(\mathcal{E}+\mathcal{P})\Theta-\frac{k^{2}\delta q}{a^{2}}+\dot{q}\Theta\,.\label{eq:XiDef}
\end{equation}
$\Xi$ is indeed proportional to the velocity divergence in the rest
frame, $\Theta_{\text{LL}}$ as defined in eq. (\ref{eq:ThetaLL_def}).
However, since it is only an algebraic redefinition for us, the meaning
of the perturbation variables remains: $\delta\mathcal{E}$, $\delta\mathcal{P}$,
etc. are still those measured in the scalar frame and \emph{not} in
the rest frame. Importantly, $\Xi$ is not singular when $w=-1$.
This definition allows us to transform perturbation equations (\ref{eq:EnCons_sc})
and (\ref{eq:MOmCons_sc}) into a simpler set,
\begin{alignat}{1}
\dot{\delta\mathcal{E}} & +3H(\delta\mathcal{E}+\delta\mathcal{P})+\Xi+(\mathcal{E}+\mathcal{P})\dot{\Phi}=0\,,\label{eq:EnCons_Xi}\\
\dot{\Xi} & +5H\Xi-(\mathcal{E}+\mathcal{P})\frac{k^{2}\Psi}{a^{2}}-\frac{k^{2}}{a^{2}}\delta\mathcal{P}-\frac{2}{3}\frac{k^{4}}{a^{4}}\delta\pi=0\,.\label{eq:MomCons_Xi}
\end{alignat}
The above then need to be furnished with Einstein's equations. We
obtain the independent equations by projecting out particular parts
of the Einstein tensor and then perturbing, \textbf{
\begin{alignat}{1}
\delta(G_{\mu\nu}u^{\mu}u^{\nu})= & 2\frac{k^{2}\Phi}{a^{2}}+6H\dot{\Phi}-6H^{2}\Psi=\delta\mathcal{E}\,,\label{eq:EinPoisson}\\
 & 2\frac{k^{2}}{a^{2}}\left(\dot{\Phi}-H\Psi\right)=-\Xi,\label{eq:EinFlow}\\
 & \Phi+\Psi=\delta\pi\,,\label{eq:EinAniso}
\end{alignat}
}where eq. (\ref{eq:EinFlow}) makes the meaning of $\Xi$ as the
total energy flow explicit. Contributions from external matter would
need to be added to both the sets of equations above.

In order to solve for the evolution of the energy-density perturbations,
we need two extra ingredients: the closure relations between $\delta\mathcal{P}$,
$\delta\pi$ and $\delta\mathcal{E}$. We shall first discuss these
relations in perfect-fluid hydrodynamics\emph{ }and compare them with
k\emph{-essence} theories. We show that even for a canonical scalar
theory with a potential the standard relation between $\delta\mathcal{P}$
and $\delta\mathcal{E}$ is \emph{not }hydrodynamical, despite the
EMT having perfect-fluid form. This departure from perfect-fluid hydrodynamics
becomes only worse for more-general scalar-field models. Since all
EMTs obey equations (\ref{eq:EnCons_Xi}) and (\ref{eq:MomCons_Xi})
at linear level, the closure relations provide the totality of information
that can be accessed from gravitational interactions at the linear
level.

On the other hand, since we are investigating theories which are described
by classical equations of motion, it is possible to obtain the exact
closure relations. As we are dealing with equations of motion for
a scalar field of second order, we can always express the closure
relations in terms of variables defined on a constant-time slice.
In section \ref{sub:GenPerts}, we will present a general parameterisation
which will serve us as a book-keeping device to ease these calculations
for particular classes of models. In section \ref{sec:BDK} we will
take a specific model and perform this calculation explicitly. Our
aim will be to obtain closure relations that to the largest extent
possible depend on a set of physical parameters of the theory, such
as the equation of state or the speed of sound.

\subsection{Perfect fluids and hydrodynamics\label{sub:hydro}}

In this section we will review the properties of perfect fluids. For
a more detailed discussion, we recommend the review \cite{Andersson:2006nr}.
The perfect-fluid EMT is obtained by averaging a distribution of particles
to obtain fluid elements and takes the form
\begin{equation}
T_{\mu\nu}=\mathcal{E}u_{\mu}u_{\nu}+\mathcal{P}\perp_{\mu\nu}\,,\label{eq:PerfHydro}
\end{equation}
with the energy density and pressure the only fluid variables. In
Newtonian hydrodynamics one starts from the axioms of energy and momentum
conservation, with the separate requirement that mass be conserved.
From these the conservation of the EMT is obtained. This has to be
appropriately generalised for relativistic fluids. The energy and
momentum conservation equations become the continuity and Euler equations,
while mass is no longer conserved. This axiom needs to be generalised
to some appropriate charge conservation. 

The velocity field $u_{\mu}$ defines the rest frame (or the lagrangian
observer) for the fluid elements. This velocity must be by definition
time-like since in the construction of the EMT it is the tangent to
the world-line of individual fluid elements. It is of course possible
to observe this EMT in a different frame. However, it will then not
in general have a perfect-fluid form and the measured energy density
and pressure will be appropriately transformed. We discuss the issue
of frame choice in detail in appendix \ref{sub:boosts}. In the rest
frame, we can relate the pressure to the energy density through\textbf{
\begin{equation}
\mathcal{P}=\mathcal{P}(\mathcal{E},S)\,,\label{eq:PES}
\end{equation}
}where $S$ is the entropy. In cosmology one often considers either
adiabatic or entropy perturbations. The former are such perturbations
of the configuration of the fluid that the pressure and energy-density
perturbations are proportional to each other. The latter are such
that despite a non-vanishing pressure perturbation, the energy density
is not affected, i.e. these are perturbations only of composition
of the fluid assuming constant energy density. The perturbations can
be written as
\begin{equation}
\delta\mathcal{P}=c_{\text{s}}^{2}\delta\mathcal{E}+\sigma\delta S\label{eq:PES-pert}
\end{equation}
where we have \emph{defined }the speed of sound $c_{\text{s}}^{2}$
as the coefficient in this equation. This sound speed then determines
such physical properties as the Jeans length. As we have shown in
appendix \ref{sub:boosts}, linear frame transformations affect the
fluid variables at second order and therefore this form is independent
of the choice of frame perturbation. 

For the purpose of calculating the anisotropies in the cosmic microwave
background (``CMB''), when the photons and baryons are strongly
coupled and in equilibrium, one can think of the photon-baryon mixture
as making up one fluid. The entropy perturbations have a clear meaning
of being a perturbation in the relative densities of photons and baryons
while keeping the energy density constant. In such a case, the expression
(\ref{eq:PES-pert}) is easily calculable \cite[pg. 307]{MukhanovBook}.
This language is useful since a proof exists that at very large scales,
the entropy perturbations are suppressed and only the adiabatic ones
contribute \cite{Lyth:2004gb,Malik:2004tf,Gao:2011mz}. However, in
late-time cosmology, we are performing observations inside the Hubble
horizon and therefore the entropy perturbations cannot in general
be neglected. 

As we will show here, this description is not so helpful for classical
scalar-field theories when applied to the late universe. The issue
boils down to the question of what the degrees of freedom that are
being perturbed in order to obtain the expression (\ref{eq:PES-pert})
are. There exists another description of fluids, much closer to our
classical-scalar-field case, in terms of the thermodynamical potentials,
temperature $T$ and chemical potential $\mu$. For a hydrodynamical
perfect fluid, it is always possible to rewrite the system (\ref{eq:PES})
in terms of these two potentials,
\begin{eqnarray}
\mathcal{P} & = & \mathcal{P}(T,\mu)\,,\label{eq:PTmu}\\
\mathcal{E} & = & \mathcal{E}(T,\mu)\,.\nonumber 
\end{eqnarray}
The pressure and energy-density perturbations are now completely determined
by the perturbations in the values of $T$ and $\mu$. In general,
we now have two types of propagating modes, one corresponding to temperature
perturbations, the other to chemical-potential perturbations, i.e.
effectively composition. 

Let us for the moment simplify and assume that the fluid be adiabatic:
such a fluid is dependent on just one thermodynamical potential, e.g.~the
temperature, $\mathcal{P}(T),\mathcal{E}(T)$. For the avoidance of
ambiguity, let us also specify that we are observing this fluid in
the rest frame. The perturbations of energy density and pressure in
the rest frame are then related by 
\begin{equation}
\delta\mathcal{P}=c_{\text{s}}^{2}\delta\mathcal{E}\,,\label{eq:AdiabdP}
\end{equation}
where $c_{\text{s}}^{2}\equiv\mathcal{P}_{,T}/\mathcal{E}_{,T}$ must
have the same meaning as the sound speed defined in (\ref{eq:PES-pert}).
In this language, the temperature is the degree of freedom which determines
everything about the configuration of the fluid. Indeed, when discussing
the photon gas, one does tend to talk of temperature perturbations
rather than energy-density perturbations. For a photon fluid the chemical
potential vanishes, since photons carry no charge, and thus everything
depends just on the temperature. Perturbing the temperature perturbs
the Bose-Einstein occupation function and therefore the number density
of photons, the energy density and pressure. From this discussion,
it should be clear that the evolution of the total pressure and energy
density with respect to a time coordinate in the frame of the observer
$\tau$ is also just related by the same sound speed, 
\[
\dot{\mathcal{P}}=c_{\text{s}}^{2}\dot{\mathcal{E}}\,,
\]
since everything is just a function of $T$ and that $c_{\text{s}}^{2}=w$,
the equation-of-state parameter. 

In the non-adiabatic case, when both the temperature and the chemical
potential are relevant, there exists another direction in which the
system could fluctuate. The pressure perturbation can now be written
down as 
\begin{eqnarray*}
\delta\mathcal{P} & = & \mathcal{P}_{,T}\delta T+\mathcal{P}_{,\mu}\delta\mu\,,\\
\delta\mathcal{E}_{\text{}} & = & \mathcal{E}_{,T}\delta T+\mathcal{E}_{,\mu}\delta\mu\,.
\end{eqnarray*}
We can also write down a very similar expression for the time evolution
of the total quantities 
\begin{eqnarray*}
\dot{\mathcal{P}} & = & \mathcal{P}_{,T}\dot{T}+\mathcal{P}_{,\mu}\dot{\mu}\,,\\
\mathcal{\dot{E}} & = & \mathcal{E}_{,T}\dot{T}+\mathcal{E}_{,\mu}\dot{\mu}\,.
\end{eqnarray*}
We can eliminate the $\delta T$ perturbation by combining these two
sets of equations to obtain 
\begin{equation}
\delta\mathcal{P}=C^{2}\delta\mathcal{E}+(\dot{\mathcal{P}}-C^{2}\dot{\mathcal{E}})\frac{\delta\mu}{\dot{\mu}}\,,\label{eq:FluidDP}
\end{equation}
with $C^{2}\equiv\mathcal{P}_{,T}/\mathcal{E}_{,T}$. We can now compare
this expression to eq. (\ref{eq:PES-pert}) and see that they have
the same form. We can read off the definition of the sound speed in
terms of the thermodynamic potentials and find an expression for the
entropy perturbation, in terms of $\delta\mu$. Note that we could
have eliminated the chemical potential instead of the temperature
perturbation. It is important to stress that the form of the expression
(\ref{eq:FluidDP}) is independent of the frame of the observer, provided
that it differs from the rest frame only at linear order. In particular
it does not rely on coordinates, but is purely determined by the fluid
configuration.

The idea of this rewriting of the system (\ref{eq:PES}) in terms
of the thermodynamical potentials is that we can rewrite the system
in terms of the degrees of freedom that we are free to perturb independently
on any spatial hypersurface and then obtain an effective equation
(\ref{eq:FluidDP}) eliminating some of them. \label{dofperts}As
we will see, in a classical scalar-field theory, we also have such
underlying degrees of freedom (the field value and the magnitude of
its first derivative) on which all the fluid variables depend. The
EMTs of the scalar systems look more similar to eq.~(\ref{eq:PTmu})
than to eq.~(\ref{eq:PES}). \\

A fundamental assumption that we have used in the above is that no
hydrodynamic quantity is an explicit function of time: pressure and
energy both depend \emph{only} on the thermodynamical potentials.
The potentials evolve according to the laws of thermodynamics, changing
the configuration of the fluid, but if the potentials were to be returned
to the same values as at some point in the past, the pressure and
energy density would also \emph{return }to their previous values.
This is why the phase diagram of water does not change over time.

One could imagine violating this principle. This would really mean
that what the fluid is changes as a function of time, not just that
its configuration evolves. Let us go back to a single thermodynamic
potential, $T,$ but this time together with an explicit dependence
on rest-frame time $\tau$. We can write the derivative w.r.t. the
rest-frame time $\tau$,%
\footnote{Note that using any other time coordinate here would mix the meaning
of what should be considered to be a perturbation of the fluid in
the rest frame and a perturbation of the rest frame itself. We should
always rewrite all the variables to be in terms of rest-frame quantities
where the fluid has a perfect form.%
} 
\begin{eqnarray}
\frac{\mathrm{d\mathcal{P}}}{\mathrm{d}\tau} & = & \partial_{\tau}\mathcal{P}+\mathcal{P}_{,T}\frac{\mathrm{d}T}{\mathrm{d}\tau}\,,\label{eq:EvolFluid}\\
\frac{\mathrm{d}\mathcal{E}}{\mathrm{d}\tau} & = & \partial_{\tau}\mathcal{E}+\mathcal{E}_{,T}\frac{\mathrm{d}T}{\mathrm{d}\tau}\,.\nonumber 
\end{eqnarray}
At a single point in time, i.e. on some spatial hypersurface in the
rest frame, the perturbations will look as they do for the adiabatic
fluid, since the only perturbed variable will be $T$,
\begin{eqnarray}
\delta\mathcal{P}_{\text{rf}} & = & \mathcal{P}_{,T}\delta T\,,\label{eq:rf-pert}\\
\delta\mathcal{E}_{\text{rf}} & = & \mathcal{E}_{,T}\delta T\,.\nonumber 
\end{eqnarray}
However now, when we transform into the frame of a different observer,
the value of the perturbation observed by them is going to be corrected
by the fact that the rest-frame time and the proper-time of the two
observers are no longer the same, 
\begin{eqnarray}
\delta\mathcal{P} & = & \partial_{\tau}\mathcal{P}\delta\tau+\delta\mathcal{P}_{\text{rf}}=\mathcal{P}_{,T}\left(\delta T-\dot{T}\delta\tau\right)+\dot{\mathcal{P}}\delta\tau\,,\\
\delta\mathcal{E} & = & \partial_{\tau}\mathcal{E}\delta\tau+\delta\mathcal{E}_{\text{rf}}=\mathcal{E}_{,T}\left(\delta T-\dot{T}\delta\tau\right)+\dot{\mathcal{E}}\delta\tau\,,\nonumber 
\end{eqnarray}
where we have eliminated the partial derivatives w.r.t.~$\tau$ using
eqs (\ref{eq:EvolFluid}). Using the definition of the sound speed
in the rest frame, as we had defined it for the adiabatic fluid, we
can combine the above two results to obtain,
\[
\delta\mathcal{P}=c_{\text{s}}^{2}\delta\mathcal{E}+\left(\dot{\mathcal{P}}-c_{\text{s}}^{2}\dot{\mathcal{E}}\right)\delta\tau\,.
\]
This relation has a very similar form to that for the non-adiabatic
fluid, eq.~(\ref{eq:FluidDP}), which was defined as being dependent
on at least two thermodynamic potentials and with no explicit dependence
on time. However, its meaning is \emph{completely} different: here
the substance depends only on \emph{one }thermodynamic potential,
but the way it depends on it evolves as a function of the rest-frame
time $\tau$. Thus it is not the same fluid at two different times.
Usual hydrodynamics does \emph{not }apply in these circumstances.
Or more bluntly put: the phase diagram of such substances would not
be the same at different points in time.

If we now assume that at the background level the rest frame and the
observer's frame are the same, the transformation between the two
frames is small. The relative perturbations of coordinates are linear.
Since $\tau$ is the proper time for the rest frame, we can think
of the velocity of the rest frame as the derivative
\[
U_{\mu}=-\partial_{\mu}\tau\,,
\]
with the normalisation automatically taken care of. We can thus arrive
at the standard expression \cite{Kunz:2006wc} for the pressure perturbation
of this substance
\begin{equation}
\delta\mathcal{P}=c_{\text{s}}^{2}\delta\mathcal{E}+\dot{\mathcal{E}}\left(c_{\text{a}}^{2}-c_{\text{s}}^{2}\right)\frac{V}{k^{2}}\,,\label{eq:EvolvSubst}
\end{equation}
where we have defined the ``adiabatic sound speed'' $c_{a}^{2}$
as $\dot{\mathcal{P}}/\dot{\mathcal{E}}$ and $V=\partial_{i}\delta U^{i}$
is the divergence of the velocity field perturbation. We should underline
once more: this expression is not hydrodynamical. It explicitly depends
on the coordinates of the frame in which the observations are being
performed. $V$ should \emph{not }be thought of as an entropy perturbation.

\subsection{k-\emph{essence} closure relations\label{sub:k-ess}}

We have spent quite some time laying out the hydrodynamical picture
in order to facilitate a comparison with a classical scalar-field
theory. Here we will discuss general k-\emph{essence, }defined by
the Lagrangian 
\[
\mathcal{L}=K(\phi,X)\,.
\]
In the scalar-field variables, the EMT can be written as 
\begin{equation}
T_{\mu\nu}=\left(mK_{m}-K\right)\frac{\nabla_{\mu}\phi\nabla_{\nu}\phi}{m^{2}}+K\left(g_{\mu\nu}+\frac{\nabla_{\mu}\phi\nabla_{\nu}\phi}{m^{2}}\right)\,,\label{eq:k-ess-emt}
\end{equation}
where we are using the definition $m\equiv\sqrt{2X}$. We've written
the EMT in such a way that allows us to easily see that it takes a
perfect-fluid form when we identify the gradient of the scalar with
the rest-frame velocity, i.e. $u_{\mu}=\nabla_{\mu}\phi/m$. As we
have already stated in section \ref{sec:FluidFormalism}, this choice
identifies $\phi$ with a comoving clock for the scalar system and
therefore $m=u^{\mu}\nabla_{\mu}\phi$ with the rate of flow of this
time. There is no other choice of frame we could have made for this
model, since only with this choice of $u_{\mu}$ is the EMT of perfect-fluid
form. The EMTs of more general scalar-field models cannot be rewritten
in perfect-fluid form at all \cite{Pujolas:2011he}.

It is important to stress that this perfect-fluid description does
\emph{not }always apply for the EMT (\ref{eq:k-ess-emt}). As we have
stated in section \ref{sub:hydro} the velocity $u_{\mu}$ needs to
be time-like so that it can represent the velocity of fluid elements.
A configuration of a scalar field can in general have non-time-like
gradients (for example, a static domain wall). There is nothing wrong
with such a configuration: its evolution is described by the equation
of motion for the scalar field in the usual way. However, the EMT
for such can \emph{not }be interpreted as one of a perfect fluid.
Similarly, the EMT of an oscillating scalar configuration is not of
perfect-fluid form at all times. The velocity $\dot{\phi}$ vanishes
and reverses direction. This means that at the rate of flow of time
vanishes and reverses and the scalar reaches the maximal amplitude.
In between two maxima, we can interpret the scalar as a perfect fluid,
but this interpretation breaks down at the maxima themselves. Of course,
it is well known that an oscillating scalar field (for example, at
the end of inflation) effectively evolves as pressureless dust. Indeed
this is true. However, there we are making a statement about the properties
of the EMT averaged over oscillation cycles. This requires the existence
of an external clock which describes the oscillatory solutions and
therefore the existence of an external reference system which increases
monotonically. The EMT of the scalar in such a situation is not that
of a perfect fluid at all times, it is just that the particular solution
can be described by an averaged EMT which has the form of the EMT
for dust. 

Perfect-fluid evolution can always be written in terms of a non-canonical
scalar-field theory \cite{Schutz}; however, not all solutions of
k-\emph{essence} theories can be described by perfect fluids. The
reason why fluid descriptions are at all useful in cosmology is that
the symmetry of the FLRW background requires that the spatial gradients
vanish. Thus particular solutions always have time-like velocities
(apart from the oscillatory cases mentioned). Linear perturbation
on this background by definition do not change this property since
they are small. However, the question of whether the particular solution
behaves like a fluid is in principle very different from the question
of whether the scalar-field itself is a fluid. As we show here, this
is not always the case and therefore it should not be a surprise that
one does not in general obtain hydrodynamical relations between the
perturbations of pressure and energy density. We will show this explicitly
now.

For k\emph{-essence }models, both the pressure and the energy density
are functions of just two variables, $\phi$ and $m$. As we discussed
on page \pageref{dofperts}, it is natural to first write down the
perturbations in terms of these underlying independent variables,
\begin{eqnarray*}
\delta\mathcal{P} & = & \mathcal{P}_{m}\delta m+\mathcal{P}_{\phi}\delta\phi\,,\\
\delta\mathcal{E} & = & \mathcal{E}_{m}\delta m+\mathcal{E}_{\phi}\delta\phi\,.
\end{eqnarray*}
where the subscripts $m,\phi$ denote partial differentiation w.r.t.
those variables. We can then eliminate the derivatives w.r.t. $\phi$
by taking time derivatives of $\mathcal{E}$ and $\mathcal{P}$, obtaining
\begin{eqnarray}
\delta\mathcal{P} & = & \mathcal{P}_{m}\left(\delta m-\frac{\dot{m}}{m}\delta\phi\right)+\dot{\mathcal{P}}\frac{\delta\phi}{m}\,,\label{eq:dPdE-kess}\\
\delta\mathcal{E} & = & \mathcal{E}_{m}\left(\delta m-\frac{\dot{m}}{m}\delta\phi\right)+\dot{\mathcal{E}}\frac{\delta\phi}{m}\,.\nonumber 
\end{eqnarray}
Given the definition of the spatial velocity divergence $\Theta$,
eq. (\ref{eq:ThetaX_def}), we can combine the above to obtain the
k-\emph{essence} closure relation
\begin{equation}
\delta\mathcal{P}=C^{2}\delta\mathcal{E}+\dot{\mathcal{E}}\left(c_{\text{a}}^{2}-C^{2}\right)\frac{\Theta}{k^{2}}\,,\label{eq:k-ess-closure}
\end{equation}
where $C^{2}\equiv\mathcal{P}_{m}/\mathcal{E}_{m}$. This is the standard
result obtained in ref.~\cite{Christopherson:2008ry} and also matches
the result obtained in ref.~\cite{Battye:2012eu} when the parametrisation
is reduced to dark-energy models of k-\emph{essence} type. This relationship
is \emph{not }the hydrodynamical closure relation for a non-adiabatic
fluid (\ref{eq:FluidDP}), but rather it is of the type for a substance
(\ref{eq:k-ess-closure}) depending explicitly on its rest-frame time
(\ref{eq:EvolvSubst}), exactly since it depends on the perturbation
of the rest-frame represented by $\Theta$. Even assuming that the
gradients $\nabla_{\mu}\phi$ are time-like, a general k-\emph{essence}
theory is not hydrodynamical, but represents a fluid which itself
changes with time. There is only one thermodynamical potential, $m$,
and there are no entropy perturbations in this model. However, in
the shift-symmetric case, $K=K(X)$, there is no explicit $\phi$
dependence, as a result of which $c_{\text{a}}^{2}=C^{2}$. In this
case, the coordinate dependence of the closure relation (\ref{eq:k-ess-closure})
disappears and we recover the standard behaviour for a perfect adiabatic
fluid, eq.~(\ref{eq:AdiabdP}). However, this is a fluid which can
only carry scalar perturbations and not vorticity: a superfluid.%
\footnote{As we have already stated, this is a result of the Frobenius theorem.
The rotation tensor for the velocity vector field $u^{\mu}$ vanishes,
$\omega_{\mu\nu}\equiv\boldsymbol{\overline{\nabla}}_{[\mu}u_{\nu]}=0$,
since $u_{\mu}$ is a derivative of a scalar. The vorticity vector
for the vector field $u_{\mu}$ is then defined as $\omega^{\mu}\equiv\varepsilon^{\alpha\beta\gamma\mu}\omega_{\alpha\beta}u_{\gamma}$
and therefore also is zero.%
} 

One final aspect that needs to be checked is the relation of $C^{2}$
to what we call the physical sound speed $c_{\text{s}}^{2}$. $C^{2}$
relates the pressure and energy-density perturbation through eq.~(\ref{eq:k-ess-closure}),
and therefore determines the Jeans length. What we will call the \emph{physical
}sound speed $c_{\text{s}}^{2}$ in the case of a scalar-field theory
is defined as the speed of propagation of small perturbations of the
\emph{field}. This physical sound speed determines the rate at which
information propagates and therefore the causal structure of the scalar
medium \cite{Babichev:2007dw}. It is obtained by calculating the
effective metric for perturbations. We show how this can be done in
section \ref{sub:EoM-Gmn}. In general the two quantities do not have
to be equal, but, in the case of k-\emph{essence} $C^{2}=c_{\text{s}}^{2}$
for both shift-symmetric and general models. The Jeans length in k-\emph{essence}
is therefore always determined by this physical sound speed $c_{\text{s}}^{2}$,
rather than any other speed of pressure waves \cite{Christopherson:2012kw}.
This also could have been a different result and indeed it is highly
non-trivial to obtain it in more complicated theories. However, because
of the link between the Jeans term and causality, resulting from the
fact that it describes the reaction of the fluid to changes in the
configuration, it is this physical sound speed of field perturbations
that will in the end determine the behaviour of the scalar fluid at
small enough scales.

In conclusion: k-\emph{essence} models hydrodynamics of an adiabatic
fluid only in the shift-symmetric case. Away from shift symmetry,
the substance it models changes its properties as a function of time.
However, it still depends only on one thermodynamical potential and
does not carry entropy perturbations. One can see that a non-adiabatic
perfect fluid would be modelled by two shift-symmetric scalar fields.
The two magnitudes of the gradients would represent two different
thermodynamic potentials, while the independence from the field values
would ensure that the fluid were unchanging. The description of perfect-fluid
hydrodynamics through scalar theories was discussed in ref.~\cite{Schutz}.

As we will start to show in the next section, the closure relations
for more complicated models depart even further from the hydrodynamical
picture. Nonetheless, just as in the case of eq.~(\ref{eq:dPdE-kess}),
hydrodynamic quantities such as the equation of state and the sound
speed do appear in the closure relations and therefore one can hope
to express the general behaviour using a small set of physical parameters.
We shall therefore be very careful when deriving the closure relations
and will show explicitly that the correct relations are \emph{not
}those of hydrodynamics.\\

Let us invest a little time to rewrite the perturbations for k-\emph{essence}
(\ref{eq:dPdE-kess}) using a different set of variables, which will
make the transition to more difficult cases discussed in section \ref{sec:BDK}
more transparent. The pressure perturbation can be considered to be
a sum of two parts: spatial and temporal, 
\begin{equation}
\frac{k^{2}}{a^{2}}\delta\mathcal{P}=m\mathcal{P}_{m}A+\dot{\mathcal{P}}\Theta\,.\label{eq:dP-A-kess}
\end{equation}
with a similar result for the energy-density perturbations. We have
defined the variable $A$,
\begin{equation}
A\equiv\left(\frac{k^{2}}{a^{2}}\frac{\delta m}{m}-\frac{\dot{m}}{m}\Theta\right)=\dot{\Theta}+2H\Theta-\frac{k^{2}}{a^{2}}\Psi\,,\label{eq:A-def}
\end{equation}
to represent the perturbation of $m$ as it would be seen in the scalar
frame $u^{\mu}$, i.e.~it is the equivalent of the rest-frame perturbation
(\ref{eq:rf-pert}). It is also equivalent to the spatial divergence
of the perturbation of the acceleration $a_{\lambda}$ (see eq.~(\ref{eq:a_as_m})).
Spatial in this case implies independent of the perturbation of the
clock $\phi$.

We then obtain the closure relations by rewriting the velocity divergence
$\Theta$ in eq.~(\ref{eq:dP-A-kess}) through the total energy flux
$\Xi$ (eq.~(\ref{eq:XiDef})) and then eliminating $A$ from the
pressure perturbation equation using the energy-density perturbation.
We have to do this since only $\Xi$ and not $A$ has an evolution
equation (eq.~(\ref{eq:MomCons_Xi})) in the language that we have
set up. In this language the closure relations (\ref{eq:k-ess-closure})
for k-\emph{essence} become 
\begin{alignat}{2}
\text{k-\emph{essence}:} &  & \delta\mathcal{P} & =c_{\text{s}}^{2}\delta\mathcal{E}+3(c_{\text{s}}^{2}-c_{\text{a}}^{2})\left(\frac{aH}{k}\right)^{2}\frac{\Xi}{H}\,,\label{eq:k-ess-closure-Xi}\\
 &  & \frac{k^{2}}{a^{2}}\delta\pi & =0\,.
\end{alignat}
The horizon suppression factor, $aH/k$, appears here since $\Xi$
was defined in eq.~(\ref{eq:XiDef}) as containing two spatial derivatives
already and this factor cancels out these derivatives. The k-\emph{essence
}model contains just functions of $\phi$ and $m$ and does not contain
any higher-order derivative terms, therefore the perturbations themselves
cannot contain such contributions. This changes in theories containing
second derivatives of the scalar in the Lagrangian. Apart from this,
the coefficients are just functions of time, dependent purely on the
evolution of background quantities. It is only such terms that can
reasonably be hoped to be constant or to scale simply when the scalar
is following a scaling solution. Our aim in the next section is to
extract this sort of behaviour for more general models.

\subsection{General parameterisation of closure relations\label{sub:GenPerts}}

As we have already mentioned, in order to properly solve for the evolution
of energy-density perturbations in any particular model in the fluid
language, we need to calculate the appropriate closure relations for
it. At this stage it is useful to introduce a parameterisation for
these relations. This parameterisation will be based on the expectations
of the kind of terms that more general theories of a single scalar
degree of freedom yield. Our aim is not to introduce a set of parameters
that will span the space of all possible and impossible theories and
which should be constrained from the data. After all, the most general
theory of a single scalar degree of freedom is already known \cite{Horndeski:1974,Deffayet:2011gz}.
What remains is to perform calculations for these healthy theories
within some a physically meaningful framework. Our parameterisation
should be thought of as a book-keeping device which should aid in
these calculations. As we will show, the advantage of this approach,
as opposed to just solving the equations of motion for the scalar
fields, is that it allows for understanding the effect of changing
the hydrodynamical properties such as the sound speed and equation
of state, within particular classes of models, without worrying about
the exact form of the Lagrangian.

We propose that, under the assumptions we detail below, the form of
the closure relations will be contained within the structure:
\begin{eqnarray}
\delta\mathcal{P} & = & C^{2}\delta\mathcal{E}+3\Sigma_{1}\left(\frac{aH}{k}\right)^{2}\frac{\Xi}{H}+\Sigma_{2}\frac{\Xi}{H}+\beta_{\mathcal{P}}e^{-\varkappa}\delta\rho+\gamma_{\mathcal{P}}\frac{e^{-\varkappa}\rho}{H}\Theta_{\text{m}}\left(\frac{aH}{k}\right)^{2}\,,\label{eq:ClosureParam}\\
\frac{k^{2}}{a^{2}}\delta\pi & = & \Pi\delta\mathcal{E}+3\varpi_{1}\left(\frac{aH}{k}\right)^{2}\frac{\Xi}{H}+\varpi_{2}\frac{\Xi}{H}+\beta_{\pi}e^{-\varkappa}\delta\rho+\gamma_{\pi}\frac{e^{-\varkappa}\rho}{H}\Theta_{\text{m}}\left(\frac{aH}{k}\right)^{2}\,.\nonumber 
\end{eqnarray}
where $\rho$ and $\delta\rho$ are the energy density and its perturbation
of any matter external to the dark energy assumed here to behave as
dust, $\Theta_{\text{m}}$ is the velocity divergence of the external
matter, $e^{\varkappa}$ is the effective Planck mass which appears
as a result of our having eliminated $G_{\mu\nu}$ from the EMT of
the dark energy. The set of parameters which need to be \emph{calculated
}for a particular class of models are: $C^{2},\Pi,\Sigma_{,1,2},\varpi_{1,2},\beta_{\mathcal{P},\pi}$
and $\gamma_{\mathcal{P},\pi}$. 

Let us justify this structure. Firstly, it is clearly not a structure
which would be given by hydrodynamics of a perfect fluid, just as
it wasn't in the case of k\emph{-essence. }In\emph{ }general, the
EMT will contain standard k-\emph{essence} contributions, which will
result in the type of terms obtained in eq.~(\ref{eq:k-ess-closure-Xi}).
More general scalar models contain second derivatives in the EMT,
generating entries in all the components, as can be seen from the
decomposition (\ref{eq:Bmunu}). The fluid variables can for example
depend on the expansion, $\theta$ or the intrinsic curvature of the
spatial hypersurface, $^{(3)}R$. When perturbed these give
\begin{eqnarray}
\delta\theta & = & \Theta+3(\dot{\Phi}-H\Psi)\,,\label{eq:ThetaPert}\\
\delta{}^{(3)}R & = & 4\frac{k^{2}\Phi}{a^{2}}-4H\Theta\,.\nonumber 
\end{eqnarray}
The gravitational potentials here can be solved for by using the Einstein
equations (\ref{eq:EinPoisson}), at the cost of introducing a dependence
on the energy density and velocity divergence of any external matter
present in the cosmology. Thus in general models with second derivatives,
the perturbed fluid variables will not only be a function of the scalar
configuration variables $\delta\phi$ and $\delta m$, but also be
functions of the configuration of the external matter. This is the
origin of the maybe surprising parameters $\beta_{\mathcal{P},\pi}$
and $\gamma_{\mathcal{P},\pi}$ in the parameterisation (\ref{eq:ClosureParam}).
This doesn't make the system unsolvable, since there will also be
conservation equations for the external matter equivalent to eqs (\ref{eq:EnCons_Xi})
and (\ref{eq:MomCons_Xi}), but it makes these more-general models
significantly more complex. eqs (\ref{eq:ThetaPert}) show that the
dependence on $\Theta$ arising from the second-derivative terms is
unsuppressed by $k/aH$, leading to the contributions of terms $\Sigma_{2}$
and $\varpi_{2}$ to the parameterisation (\ref{eq:ClosureParam}).

Another general feature that appears in these models with second derivatives
is the presence of $\nicefrac{\mathrm{d}m}{\mathrm{d}\tau}$, the
second time derivative of the scalar field, in quantities such as
$\mathcal{P}$ (see e.g.~eq.~(\ref{eq:EMT_fluid_decomp}) or ref.~\cite{Pujolas:2011he}).
This is yet again another independent variable on which the pressure
perturbation will depend. The method we propose for solving this complication
is to use the equation of motion of the scalar field to eliminate
$\nicefrac{\mathrm{d}m}{\mathrm{d}\tau}$, prior to perturbing. Since
any coupling to external matter will by definition also appear in
this equation of motion, this sort of replacement will again give
a contribution to the pressure of dark energy EMT from the external
matter. In the equation of motion this coupling can be more general
than just to the energy density of the external matter.%
\footnote{e.g. in Brans-Dicke theories, the equation of motion couples the scalar
field to the trace of the external EMT, i.e. $\rho-3p$, but it is
a different combination of pressure and energy for galileon terms.%
} We will neglect the pressure of the external matter in this paper
and thus have not included it in eqs (\ref{eq:ClosureParam}). 

In section \ref{sec:BDK} we illustrate the above somewhat abstract
arguments with a concrete example model. Even this simplest example
will require most of the structure of the parameterisation (\ref{eq:ClosureParam}).
We will see that theories with second-derivatives in the EMT feature
a new scale for perturbations: it is determined by the relative sizes
of the k\emph{-essence} and second-derivative terms in the perturbed
EMT. At large enough scales, the k-\emph{essence} terms will dominate
and the perturbations will evolve as they do in k-\emph{essence}.
At smaller scales, the two-derivative terms will dominate bringing
new features to the evolution of the perturbations, such as anisotropic
stress. This means that the parameters introduced in (\ref{eq:ClosureParam})
are not strictly speaking functions of just the background. They will
contain exactly this perfect/imperfect transition scale, and will
in general interpolate between two values either side of this scale.
However, the way that we have defined them, away from this transition
scale they will behave as constants in $k$ (not necessarily in time).
In particular, no higher derivative terms are possible (higher powers
of $k/aH$) if we limit ourselves to models with no more than two
derivatives. We should also note that each class of Lagrangians with
second derivatives will introduce its own transitions scale, dependent
on the value of the coefficient it carries. In this paper, we shall
only investigate a toy model with one such term.

We should remark that this transition between the two limits is in
the end determined by the relative importance of the $\Theta$ and
$\delta q$ terms in $\Xi$, eq.~(\ref{eq:XiDef}): the dominance
of the $\delta q$ terms signifies the dominance of the two-derivative
terms. It will clearly occur at small-enough scales, when the mode
$k$ is high. However, as the form of $\Xi$, eq.~(\ref{eq:XiDef}),
shows, the $q$ terms will also dominate whenever the equation of
state is close enough to that of the vacuum, $\mathcal{E}+\mathcal{P}=0$.
Thus in models which would fit the observed expansion history, having
$w\simeq-1$, the transition to the imperfect behaviour will lie at
relatively large scales even when the two-derivative corrections are
relatively suppressed and do not contribute significantly to the evolution
of the background. It is interesting to note that this transition
scale is not present in the equation of motion for the scalar. It
is purely a result of the structure of the EMT. However, since it
is the EMT that determines the behaviour of the gravitational potentials,
this scale is capable of having an observable effect.

\medskip{}

We can now insert the parameterisation (\ref{eq:ClosureParam}) into
the perturbation equations (\ref{eq:EnCons_Xi}) and (\ref{eq:MomCons_Xi})
to obtain
\begin{alignat}{1}
\dot{\delta\mathcal{E}} & +3H(1+C^{2})\delta\mathcal{E}+\left(1+3\Sigma_{2}+9\Sigma_{1}\left(\frac{aH}{k}\right)^{2}\right)\Xi=\label{eq:EnConsPara}\\
 & =S_{1}-3H\beta_{\mathcal{P}}e^{-\varkappa}\delta\rho-3\gamma_{\mathcal{P}}e^{-\varkappa}\rho\Theta_{\text{m}}\left(\frac{aH}{k}\right)^{2}\nonumber \\
\dot{\Xi} & +\left(5-3\left(\Sigma_{1}+\frac{2}{3}\varpi_{1}\right)-\frac{\mathcal{E}+\mathcal{P}}{H^{2}}\varpi_{2}-\left(\frac{k}{aH}\right)^{2}\left(\Sigma_{2}+\frac{2}{3}\varpi_{2}\right)\right)H\Xi-\label{eq:MomConsPara}\\
 & -(\mathcal{E}+\mathcal{P})\left(\Pi-\frac{1}{2}\right)\delta\mathcal{E}-\left(C^{2}+\frac{2}{3}\Pi\right)\frac{k^{2}}{a^{2}}\delta\mathcal{E}=\nonumber \\
 & =S_{2}+\left(\frac{k^{2}}{a^{2}}\left(\beta_{\mathcal{P}}+\frac{2}{3}\beta_{\pi}\right)-\frac{(\mathcal{E}+\mathcal{P})}{2}\right)e^{-\varkappa}\delta\rho+H(\gamma_{\mathcal{P}}+\frac{2}{3}\gamma_{\pi})e^{-\varkappa}\rho\Theta_{\text{m}}\nonumber 
\end{alignat}
where we have explicitly included the non-conservation terms $S_{1,2}$
which appear as a result of the evolving Planck mass in the presence
of external matter, eq. ~(\ref{eq:EMTCons}), 
\begin{eqnarray}
S_{1} & = & \delta\left(e^{-\varkappa}T_{\mu\nu}^{\text{ext}}u^{\mu}\nabla^{\nu}\varkappa\right)\,,\label{eq:NonConsPert}\\
S_{2} & = & -ik^{i}\delta\left(e^{-\varkappa}T_{\mu\nu}^{\text{ext}}\nabla^{\mu}\varkappa\perp_{\lambda}^{\nu}\delta_{i}^{\lambda}\right)\nonumber 
\end{eqnarray}
where the $\delta$ signifies that the term inside the parentheses
must be perturbed. We have also dropped some terms involving the gravitational
potential, subdominant subhorizon (i.e. negligible on scales\textbf{
$k\gg aH$},\textbf{ }which are the most interesting ones from the
observational point of view). Given this form, we can combine the
two equations to eliminate the variable $\Xi$ and obtain a single
second-order differential equation for $\delta\mathcal{E}$ with sources
related to the presence of external matter. Let us make some simplifications
first:
\begin{itemize}
\item The term in eq.~(\ref{eq:MomConsPara}) containing $\Sigma_{2}+\frac{2}{3}\varpi_{2}$
will contribute to the friction term in the evolution equation for
$\delta\mathcal{E}$. When rewritten in $\ln a$ as the time coordinate,
it can be easily seen that this will be a contribution exponentially
growing in time, large at small scales. Depending on the sign, it
either is an anti-friction term leading to small-scale instability
or a very quickly growing friction term, destroying any perturbation
growth. In the model we describe in (\ref{sec:BDK}), it is never
important and vanishes on small scales, and we will set it to zero
in what follows.
\item Since the pressure and anisotropy perturbations contain the external
matter perturbation, $\delta\rho$, the Jeans term in eq.~(\ref{eq:MomCons_Xi})
gives a scale-dependent source for the DE perturbation. We can estimate
from eq.~(\ref{eq:MomConsPara}) that, at small enough scales, the
\emph{particular solution }of the differential equation for $\delta\mathcal{E}$
we will approximately have is
\[
\delta\mathcal{E}\simeq\frac{\beta_{\mathcal{P}}+\frac{2}{3}\beta_{\pi}}{C^{2}+\frac{2}{3}\Pi}\delta\rho\,\qquad\text{when }\beta_{\mathcal{P}}+\frac{2}{3}\beta_{\pi},C^{2}+\frac{2}{3}\Pi\gg\left(\frac{aH}{k}\right)^{2}\,.
\]
This will be the solution \emph{provided} that the homogeneous modes
decay quickly enough. This is usually referred to as the \emph{quasi-static
}limit. Since\emph{ }in this configuration the dark-energy energy
density tracks the dark matter one, this would be interpreted as a
modification of the effective Planck constant in the Poisson equation,
or a fifth force, and is the sort of solution taken as the $f(R)$
modified-gravity regime. The rather interesting observation is that
in models with a DE/DM coupling, it is on scales \emph{smaller} than
the Jeans length that we observe such effective modifications of the
Newton's constant. It is \emph{below }the Jeans length that the dark
energy is clustered in models where it is coupled to the DM. However,
we must be careful as to whether this quasi-static solution neglecting
the solutions to the homogeneous equations is in fact dominant. This
may not always be the case.%
\footnote{This is, at least in spirit, similar to the ``adiabatic instability''
in coupled models \cite{Amendola:2001rc,Koivisto:2005nr,Afshordi:2005ym,Bean:2007ny}.
This instability is a result of assuming that the scalar field follows
the minimum of an effective potential adiabatically, i.e. its evolution
is quasi-static and is determined by the density perturbation of dark
matter. However, there can exist circumstances when that is not the
solution which the scalar obeys which is signified by a negative effective
sound speed squared of the combined DM/DE fluid.%
}
\end{itemize}
Neglecting the external energy density for the moment, we can obtain
an evolution equation for the energy-density perturbation of the dark
energy when it is isolated:
\begin{align}
\ddot{\delta\mathcal{E}} & +H\left(8+3\left(C^{2}-\Sigma_{1}-\frac{2}{3}\varpi_{1}\right)+\frac{2\dot{H}}{H^{2}}\varpi_{2}\right)\dot{\delta\mathcal{E}}+\label{eq:dEEvol}\\
 & +\left(1+3\Sigma_{2}\right)\left(C^{2}+\frac{2}{3}\Pi\right)\frac{k^{2}}{a^{2}}\delta\mathcal{E}+H^{2}\left[\frac{\dot{H}}{H^{2}}\left(4(1-\Pi)-3\Sigma_{2}\left(1-6\Pi\right)\right)+\right.\nonumber \\
 & \left.+\left(3-2\Pi\right)\left(5+\frac{2\dot{H}}{H^{2}}\varpi_{2}-3\left(\Sigma_{1}+\frac{2}{3}\varpi_{1}\right)\right)\right]\delta\mathcal{E}=0\nonumber 
\end{align}
where we have dropped terms suppressed below the horizon and assumed
$\Sigma_{2}$ and $C^{2}$ vary slowly.

At this stage let us stress that it is the combination of parameters
$(1+3\Sigma_{2})(C^{2}+\nicefrac{2}{3}\Pi)$ that determines the Jeans
length. As we will see, in our example model, it is this \emph{combination
}that will reduce to the physical sound speed squared, $c_{\text{s}}^{2}$
at small enough scales. Thus both the Jeans length and the speed of
propagation of perturbations of the scalar will be determined by the
same physical sound speed rather than a simple parameter relating
$\delta\mathcal{P}$ and $\delta\mathcal{E}$. 

We would also like to draw the reader's attention to the coefficient
of the friction term in eq.~(\ref{eq:dEEvol}). It contains corrections
$\varpi_{1,2}$ which appear whenever the anisotropic stress is present.
When the fluid is imperfect, the friction coefficient is changed (in
our example it is reduced) and this modifies the power law with which
the homogeneous modes of eq.~(\ref{eq:dEEvol}) evolve, making $\delta\mathcal{\ensuremath{E}}$
decay more slowly (and the density contrast grow significantly faster)
compared to its behaviour in a perfect regime. In hydrodynamics, viscous
stresses act to increase the friction, so this result is very much
against that intuition, underlining the non-hydrodynamical behaviour
of these scalar models.

We will now turn to a discussion of a particular model that features
second derivatives in the EMT within the framework proposed in this
general discussion.

\section{Worked example: non-minimally coupled k-\emph{essence}\label{sec:BDK}}

The addition of higher-derivative terms to effective actions for cosmological
perturbations was shown to stabilise them on NEC-violating backgrounds
already in Refs~\cite{Creminelli:2006xe,Creminelli:2008wc}. But
it was the (re-)discovery of galileon-type theories \cite{Nicolis:2008in,Nicolis:2009qm,Horndeski:1974,Deffayet:2009wt,Deffayet:2009mn,Deffayet:2011gz}
that allowed for a complete modelling of the background and perturbations
together. A new class of explicitly stable NEC-violating attractor
solutions different from k-\emph{essence,} where these higher-derivative
terms are large in the background, was found in ref.~\cite{Deffayet:2010qz,Kobayashi:2010cm}.
The fundamental difference as to more phenomenological approaches,
is that these theories do have a well-defined Lagrangian and equations
of motion which allow for a precise understanding of the evolution.
Cosmological perturbation theory in these models has already been
extensively studied \cite{Kobayashi:2009wr,Kobayashi:2010wa,DeFelice:2010as,DeFelice:2010gb,DeFelice:2010jn,DeFelice:2010nf,DeFelice:2010pv,Kobayashi:2011pc,DeFelice:2011zh,Appleby:2011aa,Appleby:2012ba,Barreira:2012kk}.
However, the presence of second-order derivatives in the EMT complicates
matters very significantly, obscuring the physical meaning of the
results. 

The purpose of this paper is to show how one can use intuitive hydrodynamical
language to study these more complicated theories. However, in the
interest of simplicity and familiarity, we will study a much simpler
model than the full Horndeski-type Lagrangian \cite{Horndeski:1974}:
a k-\emph{essence }theory non-minimally coupled to gravity. Again,
this class of models has been studied before \cite{Amendola:2004qb,Wei:2004rw},
however from a very different point of view. We will focus on the
fact that in such theories second derivatives in the EMT are generated
through the non-minimal coupling to gravity. As a result of this,
in the Jordan frame, this scalar theory can be seen to exhibit many
of the properties of general galileon EMTs at the level of linear
perturbations on a FLRW background. As we will see, this theory also
features a rest frame different to the scalar frame as well as non-zero
viscous stress thus addressing most of the possible structures in
the more general theories. This worked example will allow us to demonstrate
explicitly how one goes about deriving the parameters of the closure
relations (\ref{eq:ClosureParam}), resulting in a description which
depends on physical properties of the fluid rather than a particular
choice of Lagrangian to which the equation of motion is sensitive.

As we have already mentioned in section \ref{sub:EvolEqLin}, these
models are usually described in the Einstein frame, where the coupling
to gravity is removed at the price of a coupling between external
matter and the scalar. In the Einstein frame, the second-derivative
terms in the EMT are not present and the DE fluid is just a standard
k\emph{-essence }coupled to external matter\emph{. }We\emph{ }choose
to remain in the Jordan frame, since we are using this worked example
to illustrate the features of more-complex models, such as kinetic
gravity braiding, where no frame redefinition exists which would remove
the second-derivative terms.

\subsection{General properties of energy-momentum tensor\label{sub:genprops}}

Our worked example is described by the following action: 
\begin{equation}
S=\int\mathrm{d}^{4}x\sqrt{-g}\left[\frac{1}{2}M_{\text{Pl}}^{2}e^{\varkappa(\phi)}R+K(\phi,X)+\mathcal{L}_{\text{ext}}\right].\label{eq:action1}
\end{equation}
The first term defines the gravitational part of the action, which
is the usual Einstein-Hilbert term but non-minimally coupled to a
scalar $\phi$ through the function $\varkappa(\phi)$. The second
term, $K(\phi,X)$, is the usual k-\emph{essence} term: a non-canonical
kinetic term for the scalar field. Further, through $\mathcal{L}_{\text{ext}}$
we have indicated the possibility of some matter content external
to the scalar field, which we do not consider in detail in this paper.
$M_{\text{Pl}}$ denotes the fundamental Planck mass, implying that
the effective Newton's constant is
\[
G_{\text{eff}}=G_{\text{N}}e^{-\varkappa(\phi)}\,.
\]
We will use units in which $M_{\text{Pl}}^{2}=(8\pi G_{\text{N}})^{-1}=1$.
The always-positive form of the non-minimal coupling, $e^{\varkappa}$,
ensures that the effective Newton's constant does not change sign.
If we start the system off in a configuration where the gravitational
sector does not have a ghost, it will never evolve to destabilise
it. 

We will simplify the notation somewhat by using the overdot to signify
the differentiation w.r.t. the scalar-frame time, $\dot{f}\equiv\mathrm{d}f/\mathrm{d\tau}$
in section \ref{sub:genprops} and \ref{sub:EoM-Gmn}. Once we go
back to the discussion of perturbations in section \ref{sub:DEonly_perts},
the overdot will again mean the differentiation w.r.t. background
coordinate time in FLRW.

The Einstein equations are
\begin{align}
G_{\mu\nu} & =T_{\mu\nu}^{X}+e^{-\varkappa}T_{\mu\nu}^{\text{ext}}\,,\label{eq:Einstein}\\
T_{\mu\nu}^{X} & \equiv e^{-\varkappa}K_{,X}\nabla_{\mu}\phi\nabla_{\nu}\phi+e^{-\varkappa}Kg_{\mu\nu}+e^{-\varkappa}\left(\nabla_{\mu}\nabla_{\nu}-g_{\mu\nu}\Box\right)e^{\varkappa(\phi)}\,.\label{eq:TX}
\end{align}
where $T_{\mu\nu}^{X}$ is what we will refer to as the EMT of the
dark energy and which includes the fact that we have divided the equations
through by the effective Planck mass (see the discussion on page \pageref{eq:Einstein!+f}).
$T_{\mu\nu}^{\text{ext}}$ is EMT for any external matter. If we consider
it, we will take it to be a perfect fluid,
\begin{equation}
T_{\mu\nu}^{\text{ext}}=\left(\rho_{\text{ext}}+p_{\text{ext}}\right)v_{\mu}v_{\nu}+p_{\text{ext}}g_{\mu\nu}\,,\label{eq:extEMT}
\end{equation}
and we will assume that it be conserved, $\nabla^{\mu}T_{\mu\nu}^{\text{ext}}=0$.
As was alluded to in the introduction to section \ref{sec:BDK}, the
non-minimal coupling to the Ricci scalar in the action (\ref{eq:action1})
results in second derivatives appearing in the energy momentum tensor
eq. (\ref{eq:TX}). In this respect this model shares many features
with the more complex theories featuring galileon terms or kinetic
gravity braiding from the perspective of cosmological perturbation
theory. The price of the varying Planck mass is the non-conservation
of our choice of the DE EMT in the presence of external matter,
\begin{equation}
\nabla^{\mu}T_{\mu\nu}^{X}=-\dot{\varkappa}e^{-\varkappa}T_{\mu\nu}^{\text{ext}}u^{\mu}\,,\label{eq:BDKnoncons}
\end{equation}
where the velocity $u^{\mu}$ is that defined in eq. (\ref{eq:u_def}).
and
\[
\dot{\varkappa}=m\varkappa_{\phi}\,,
\]
where the subscript $\phi$ represents differentiation with respect
to that variable. We can now decompose the dark-energy EMT in the
scalar frame (\ref{eq:u_def}) according to (\ref{eq:Tmunu_decomp}):
\begin{alignat}{2}
\mathcal{E} & =e^{-\varkappa}E-\dot{\varkappa}\theta\,, & E & \equiv mK_{m}-K\label{eq:EMT_fluid_decomp}\\
\mathcal{P} & =e^{-\varkappa}P+m^{2}(\varkappa_{\phi\phi}+\varkappa_{\phi}^{2})+\dot{\varkappa}\left(\frac{\dot{m}}{m}+\frac{2}{3}\theta\right)\,,\quad & P & \equiv K\nonumber \\
q_{\lambda} & =\dot{\varkappa}a_{\lambda}=\overline{\boldsymbol{\nabla}}_{\lambda}q\,, & q & \equiv-\dot{\varkappa}\nonumber \\
\tau_{\mu\nu} & =-\dot{\varkappa}\sigma_{\mu\nu}=\dot{\pi}\sigma_{\mu\nu}\,, & \pi & \equiv-\varkappa\nonumber 
\end{alignat}
The two variables $E$ and $P$ are the energy density and pressure
arising directly from the k-\emph{essence }term in the action, $K(\phi,X)$,
with the subscript $m$ signifying partial differentiation w.r.t.
$m$. In the spirit of eqs. (\ref{eq:potentials}) we have explicitly
provided the scalar potentials for the energy flow and viscous stress,
$q$ and $\pi$. 

The decomposition (\ref{eq:EMT_fluid_decomp}) explicitly shows up
the imperfect corrections to the EMT, all carrying two derivatives
of the scalar $\phi$. In particular, we have a contribution to the
energy density proportional to the expansion, a term in the pressure
proportional to $\dot{m}$ and a non-vanishing energy flow in the
scalar frame, proportional to the acceleration. These are the same
modifications as in the case of kinetic gravity braiding \cite{Pujolas:2011he},
which provides the motivation for us to study this otherwise well-known
model. In addition to the above corrections, in non-minimally coupled
scalar models, non-vanishing viscous stress appears, here in the form
of shear viscosity. There is also a bulk viscosity term in the pressure
proportional\textbf{ }to the expansion $\theta$.%
\footnote{Since we are starting with a Lagrangian, the system must be conservative.
It should therefore be stressed that these viscosity coefficients
are not the same as one would obtain through the gradient expansion.
They do imply that the evolution of the system depends on the shear
and expansion, similarly to the way that shear and bulk viscosity
do, but is not in fact dissipative. See the discussion in ref. \cite[\S 3.7]{Pujolas:2011he}
where such a situation is discussed in the case of the kinetic-gravity-braiding
model.%
} All these second-order terms appear with a common coefficient, $\dot{\varkappa}$,
which effectively provides a single additional parameter controlling
such modifications. The model reduces to k-\emph{essence }plus a constant
correction to the Planck mass in the limit $\varkappa=\text{const}$.

\subsection{Equation of motion and effective metric for perturbations\label{sub:EoM-Gmn}}

The equation of motion for the scalar that is obtained directly by
varying the action (\ref{eq:action1}) is
\begin{equation}
\nabla_{\mu}\left(K_{X}\nabla^{\mu}\phi\right)-K_{\phi}=\frac{1}{2}\varkappa_{\phi}e^{\varkappa}R\,.\label{eq:EoM-components}
\end{equation}
where we obtain the Ricci tensor from the Einstein equations (\ref{eq:Einstein}), 

\[
R=\mathcal{E}-3\mathcal{P}+e^{-\varkappa}\left(\rho_{{\rm ext}}-3p_{\text{ext}}\right)\,.
\]
Combining these and rewriting the second derivatives of the scalar
in terms of the kinematical decomposition (\ref{eq:Bmunu}) we obtain
for the equation of motion,

\begin{alignat}{1}
\left(E_{m}+\frac{3}{2}m\varkappa_{\phi}^{2}e^{\varkappa}\right)\dot{m}+ & m\left(P_{m}+\frac{3}{2}m\varkappa_{\phi}^{2}e^{\varkappa}\right)\theta+mE_{\phi}=\label{eq:Scalar_Eom}\\
= & \frac{1}{2}\dot{\varkappa}\left(E-3P-3m^{2}e^{\varkappa}\left(\varkappa_{\phi\phi}+\varkappa_{\phi}^{2}\right)+\rho_{\text{ext}}-3p_{\text{ext}}\right).\nonumber 
\end{alignat}
Because of the presence of the second derivatives in $\mathcal{E}$
and $\mathcal{P}$ and the non-minimal coupling, there is now a Planck-mass-suppressed
contribution which corrects the coefficients of the second-derivative
terms, $\dot{m}$ and $\theta$. Using this form we can also explicitly
see that this scalar is coupled to the trace of the external EMT.
This has an important impact on the dynamics of the scalar in the
presence of external matter.

An important feature of the form of the scalar equations of motion
(\ref{eq:Scalar_Eom}), with all the second-derivative terms explicitly
collected together, is that it allows us to calculate the sound speed
by identifying the effective metric for the propagation of linear
perturbations. The idea is that we linearise the equation of motion
around some background configuration (including gravity) and then
collect all the highest derivative terms. We will obtain an operator
$\mathcal{G_{\mu\nu}}$,
\[
\mathcal{G}^{\mu\nu}\nabla_{\mu}\nabla_{\nu}\delta\phi+\mathcal{O}\left(\nabla_{\mu}\delta\phi,\delta\phi,\Phi,\Psi\right)=0\,.
\]
Now given the standard eikonal ansatz, $\delta\phi=\mathcal{A}(x)\exp\left(i\omega\mathcal{S}(x)\right)$,
with some slowly varying amplitude $\mathcal{A}(x)$ and then taking
the formal limit $\omega\rightarrow\infty$, we obtain
\[
\mathcal{G^{\mu\nu}}\nabla_{\mu}\mathcal{S}\nabla_{\nu}\mathcal{S}=0\,.
\]
The tensor $\mathcal{G}_{\mu\nu}$ is now the effective contravariant
metric for the propagation of perturbations \cite[App. A]{Babichev:2007dw}.
In the case of our equation of motion (\ref{eq:Scalar_Eom}) on an
arbitrary background, it remains a simple diagonal metric, just as
in k-\emph{essence},
\begin{equation}
\mathcal{G_{\mu\nu}}=\left(\frac{E_{m}}{m}+\frac{3}{2}\varkappa_{\phi}^{2}e^{\varkappa}\right)u_{\mu}u_{\nu}+\left(\frac{P_{m}}{m}+\frac{3}{2}\varkappa_{\phi}^{2}e^{\varkappa}\right)\perp_{\mu\nu}\,.\label{eq:AcousticMetric}
\end{equation}
Note that this is an exact expression valid for any background metric,
not just, for example, for FLRW. We do have a correction to the k-\emph{essence
}result, with the non-minimal coupling to gravity resulting in $M_{\text{Pl}}^{2}$-suppressed
terms coming from the second-derivative terms in the DE EMT. From
this result, we can immediately read off the condition that the scalar
degree of freedom not be a ghost,%
\footnote{This is in addition to the condition $G_{\text{eff}}>0$, which is
required to keep the gravity sector healthy.%
}
\[
D\equiv\frac{E_{m}}{m}+\frac{3}{2}\varkappa_{\phi}^{2}e^{\varkappa}>0\,,
\]
which comes from the coefficient of the time-time part of the acoustic
metric. In addition, the sound speed for the propagation of small
scalar-field perturbations is the ratio of the diagonal space components
of the metric (\ref{eq:AcousticMetric}) to the time-time part, i.e.
\begin{align}
c_{\text{s}}^{2} & =\frac{P_{m}}{mD}+3\beta>0\,,\label{eq:cs2}\\
\beta(\phi,m) & \equiv\frac{\dot{\varkappa}^{2}e^{\varkappa}}{2m^{2}D}>0\nonumber 
\end{align}
where we have defined a new positive coupling \textbf{$\beta$ }arising
from the non-minimal coupling to gravity. Again, the sound speed squared
needs to remain positive in order to prevent short-timescale gradient
instabilities from developing, since this is the quantity that enters
the dispersion relation. We can rewrite (\ref{eq:cs2}) in the better-known
form,
\begin{eqnarray*}
c_{\text{s}}^{2} & = & \frac{K_{X}+\frac{3}{2}\varkappa_{\phi}^{2}e^{\varkappa}}{K_{X}+2XK_{XX}+\frac{3}{2}\varkappa_{\phi}^{2}e^{\varkappa}}\,,\\
\beta & = & \frac{\varkappa_{\phi}^{2}e^{\varkappa}}{2K_{X}+4XK_{XX}+3\varkappa_{\phi}^{2}e^{\varkappa}}\,.
\end{eqnarray*}
The sound speed is a combination of the standard expression for k\emph{-essence}
and a correction from the non-minimal coupling. The sound speed is
always equal to the speed of light for a canonical kinetic term. On
ghost-condensate-type attractors ($P_{m}=0)$ the sound speed is non-zero
as a result of the modification from the coupling to gravity. Thus
if the non-minimal coupling to gravity is weak ($\varkappa(\phi)$
is slowly varying) and we have approximate shift symmetry in the $K(\phi,X)$
term, we would expect the solution to approximately follow the k-\emph{essence}
attractor, with $w\simeq-1$ while the sound speed would naturally
be close to $3\beta$, at least when the dark energy is dominant. 

In general, the sound speed $c_{\text{s}}^{2}>3\beta$. This restriction
could perhaps be avoided for some functions $K(\phi,X)$ involving
inverse powers of $X$. However, these are likely to have additional
pathologies in their phase space and we shall not consider such models
here and assume this hierarchy. 

It should be stressed that the result (\ref{eq:cs2})\textbf{ }is
valid for the action (\ref{eq:action1})\textbf{ }and is not a general
prescription for the sound speed. See e.g.~refs \cite{Deffayet:2010qz,DeFelice:2011bh}\textbf{
}for the modifications to the expressions generated by galileon-type
terms in the Lagrangian. For the avoidance of confusion with other
uses in the literature, we will refer to eq.~(\ref{eq:cs2}) as the
\emph{physical }sound speed. The key remark here is that this metric
(\ref{eq:AcousticMetric}) describes the effective space time (and
its causal structure) that a perturbation in the scalar field sees.
Thus signals sent using the scalar field would propagate in this metric
and therefore with the speed (\ref{eq:cs2}). \\

One should ask about the relevance of the speed of propagation of
scalar signals to what is normally considered in the study of cosmological
perturbations: the sound speed is normally taken to relate the pressure
and energy-density perturbations; frequently it is even \emph{defined
}as $\delta\mathcal{P}/\delta\mathcal{E}$. One speaks about the speed
of pressure waves with the idea that the response from pressure is
what stops collapse and gives the Jeans length. As we will show, in
the model (\ref{eq:action1}) this reaction to collapse is not just
driven by the pressure $\mathcal{P}$ alone, but also by the anisotropic
stress, as was already previewed in eq.~(\ref{eq:dEEvol}). Moreover,
we will show that despite the fact that the relationship between the
pressure and energy-density perturbations for scalar-field theories
is in general complicated, the Jeans length is determined by the actual
physical sound speed (\ref{eq:cs2}). Thus the Jeans length, even
in these more complex theories is determined by causality and the
rate of propagation of signals in the scalar fluid. There is only
ever one physical sound speed and it is determined by the configuration
of the cosmological background and therefore, in FLRW cosmology, can
at most be a function of time.

As we show in section \ref{sub:DEonly_perts}, the quantity $\beta$
is associated with the existence of a new scale delimiting two behaviours
of cosmological perturbations: it separates perfect and imperfect
regimes. We will show that this is a separate scale from the Jeans
scale discussed above and does not appear in the equation of motion.
It only arises as a result of the structure of the EMT.\\

We shall now return to the discussion of perturbations in the fluid
language. As we have stated in section \ref{sub:GenPerts}, we need
to eliminate any terms in the EMT which are not defined on the spatial
hypersurface. In our model (\ref{eq:action1}), this means eliminating
$\dot{m}$ from the pressure in the EMT (\ref{eq:EMT_fluid_decomp}).
It is easy to rewrite the equation of motion (\ref{eq:Scalar_Eom})
in a form which will be directly usable:
\begin{equation}
\dot{\varkappa}\frac{\dot{m}}{m}=-\dot{\varkappa}c_{\text{s}}^{2}\theta-\frac{\dot{\varkappa}E_{\phi}}{mD}+\beta e^{-\varkappa}\left(E-3\left(P+m^{2}e^{\varkappa}(\varkappa_{\phi\phi}+\varkappa_{\phi}^{2})\right)+\rho_{\text{ext}}-3p_{\text{ext}}\right)\,.\label{eq:EoM-m.}
\end{equation}
We can now use eq. (\ref{eq:EoM-m.}) directly in eq. (\ref{eq:EMT_fluid_decomp})
to re-express the pressure as
\begin{align}
\mathcal{P}= & (1-3\beta)\left(e^{-\varkappa}P+m^{2}(\varkappa_{\phi\phi}+\varkappa_{\phi}^{2})\right)+\beta e^{-\varkappa}\left(E-\frac{2mE_{\phi}}{\dot{\varkappa}}\right)\label{eq:P-no_mdot}\\
 & -\left(c_{\text{s}}^{2}-\frac{2}{3}\right)\dot{\varkappa}\theta+\beta e^{-\varkappa}(\rho_{\text{ext}}-3p_{\text{exp}})\,.\nonumber 
\end{align}
We now see some of the structure which we previewed in section \ref{sub:GenPerts}:
the coupling of the scalar to external matter appears explicitly in
the pressure with the coupling parameter $\beta$. Secondly, a term
proportional to the physical sound speed appears as a coefficient
of the second-derivative term, $\theta$. It is this term that will
dominate the pressure at small scales and determine the Jeans scale. 

Another implication of the the pressure's containing $\dot{m}$ is
the presence of singularities in the phase space: as can be explicitly
seen in eq.~(\ref{eq:P-no_mdot}). Whenever $D\rightarrow0$, we
will also have \textbf{$\beta\rightarrow\infty$ }and so a number
of terms in the pressure will diverge. For a cosmological background
solution, if the pressure in this limit is negative, the trajectory
will evolve toward a pressure singularity: $\dot{H}$ will diverge
within a finite time, while $a$ and $H$ remain finite \cite{Barrow:2004xh}.
This sort of behaviour can also be seen in phase spaces of models
with kinetic gravity braiding in ref.~\cite{Easson:2011zy} and arises
in the same way. We do not observe this in minimally coupled k-\emph{essence},
despite the possibility of having configurations with a vanishing
$D$ since the pressure does not contain $\dot{m}$ directly.\\

Let us close this discussion of the general properties of the models
(\ref{eq:action1}) by discussing the limit to $f(R)$ gravity, which
is a subclass. The action is 
\begin{equation}
S=\int\mathrm{d}^{4}x\sqrt{-g}M_{\text{Pl}}^{2}f(R)\,.\label{eq:f(R)}
\end{equation}
The standard procedure is to start off from this action and rewrite
it in the Einstein frame, which explicitly shows that an extra scalar
degree of freedom is present. However, even in the Jordan frame one
can perform a Legendre transformation introducing a scalar explicitly
and obtaining the action 
\[
S=\int\mathrm{d}^{4}x\sqrt{-g}M_{\text{Pl}}^{2}\left[f_{\phi}(\phi)R+f(\phi)-\phi f_{\phi}(\phi)\right]\,,
\]
equivalent to (\ref{eq:f(R)}) \cite{Chiba:2003ir}. This form of
the action shows that $f(R)$ gravity is equivalent to our class of
models (\ref{eq:action1}) in the limit where the k-\emph{essence
}function $K$ is purely a potential $V(\phi)$. In the Jordan frame,
the kinetic term for the scalar is \emph{purely} generated through
the coupling to gravity. In the EMT (\ref{eq:EMT_fluid_decomp}),
apart from the potential energy $V(\phi)$, there are only terms arising
from the second derivatives and the behaviour of the fluid is always
determined solely by them. In particular, we have 
\begin{eqnarray}
\text{no scalar ghost}\quad D & > & 0\quad\Leftrightarrow\quad f_{\phi}>0\,,\label{eq:f(R)props}\\
c_{\text{s}}^{2} & = & 1,\nonumber \\
\beta & = & 1/3.\nonumber 
\end{eqnarray}
Thus $f(R)$ is a special case of the class of models (\ref{eq:action1})
where the Jeans scale and the transition scale associated with $\beta$
are unobservable, since $c_{\text{s}}^{2}$ and $\beta$ are both
order one. In cosmology, both in the background evolution and in the
perturbations, the second-derivative terms dominate the DE EMT on
all scales. One can reverse this observation and note that we approximately
recover the behaviour of $f(R)$ solutions, including the properties
(\ref{eq:f(R)props}) for any Lagrangian belonging to the whole class
(\ref{eq:action1}) whenever the background configuration is such
that the kinetic energy arising from the k-\emph{essence} term is
small compared to the contribution to the energy density arising from
the second-derivative terms, i.e. for background cosmology when 
\begin{equation}
e^{-\varkappa}\left(E+P\right)\ll\dot{\varkappa}H\,.\label{eq:deepMDE}
\end{equation}
Since $H$ contains a contribution from external matter, given everything
else be fixed, this condition is more likely to be satisfied in the
past, when dark matter dominates the universe, although whether this
happens is solution dependent. This type of solution was studied under
the name of the ``strong-coupling'' regime for Brans-Dicke theories
in ref. \cite{Amendola:1999qq}. 

When condition (\ref{eq:deepMDE}) is not satisfied implying that
the k-\emph{essence} terms dominate, the phenomenology of the theories
(\ref{eq:action1}) is much richer: both the sound speed and the coupling
$\beta$ can take small values, placing new potentially observable
scales affecting the evolution of dark-energy density perturbations
inside the cosmological horizon. In general, the history of the evolution
of such a scalar will go through both the $f(R)$-like regime and
this k\emph{-essence-}dominated regime. The history of this evolution
is highly model dependent. Since the purpose of this paper is to propose
a prescription for studying such models, we will only illustrate our
method in the somewhat unphysical case of a cosmology of an isolated
dark-energy fluid, corresponding to the far future, when the dark
matter has been diluted away by the expansion of the universe. This
will allow us to demonstrate a number of general properties for scalar
theories involving second derivatives in the EMT. However, we will
return to a more detailed study of the full richness of these models
in the presence of dark matter in a separate work.

\subsection{Cosmological perturbations\textmd{\label{sub:DEonly_perts}}}

In this section, we will study the behaviour of the non-minimally
coupled k-\emph{essence }model (\ref{eq:action1}) at the linear level
in cosmological perturbations using the prescription introduced in
section \ref{sub:GenPerts}. For simplicity, we will only consider
the case of a universe containing \emph{solely }the scalar field.
This will be sufficient to prove a number of statements we have made
so far, but we will not explore the full richness of the models in
this work. However, at appropriate moments, we will restore the dependence
on the external matter to illustrate generic properties. We will not
address the issues of non-linear perturbations and the existence of
the chameleon mechanism, which can change the behaviour of the scalar
at small-enough scales. 

We will explicitly show that this model features a new length scale
which divides the regimes of perfect and\emph{ }imperfect behaviour
of the system. In the limit of $f(R)$ theories, this length scale,
as well as the Jeans scale, lie beyond observable scales. It arises
in our fluid language as a result of the dominance in the pressure
and energy-density \emph{perturbations }of terms containing second
derivatives of the scalar field. Such terms are generic in other models
containing second derivatives, such as the galileon or kinetic-gravity
braiding, therefore we expect such a length scale and transitions
between two regimes with different perturbation behaviour to also
be generic. Secondly, we will prove that the Jeans scale is purely
dependent on the speed of propagation of the scalar-field perturbations,
the physical sound speed (\ref{eq:cs2}) and not on the naive relationship
between the pressure and energy-density perturbations. 

From here on now the overdot will signify the derivative w.r.t.~the
FLRW background coordinate time, $\mathrm{d}/\mathrm{d}t$. The prime
represents the derivative w.r.t. $\ln a$.\\

We will now explicitly demonstrate how the closure relations in our
example model reduce to the form proposed in the parameterisation
(\ref{eq:ClosureParam}). Let us start by reminding ourselves of the
expressions for the DE energy density and pressure, eqs (\ref{eq:EMT_fluid_decomp}):
\begin{align}
\mathcal{E}= & e^{-\varkappa}E-\frac{\mathrm{d}\varkappa}{\mathrm{d}\tau}\theta\,,\label{eq:E-sec3B}\\
\mathcal{P}= & e^{-\varkappa}\left((1-3\beta)\left(P+m^{2}(\varkappa_{\phi\phi}+\varkappa_{\phi}^{2})\right)+\beta\left(E-\frac{2E_{\phi}}{\mathrm{\varkappa_{\phi}}}\right)\right)-\left(c_{\text{s}}^{2}-\frac{2}{3}\right)\theta\frac{\mathrm{d}\varkappa}{\mathrm{d}\tau}\,.\label{eq:P-sec3B}
\end{align}
These are \emph{exact }covariant expressions in the absence of external
matter and where the dependence on $\mathrm{d}m/\mathrm{d}\tau$ has
been eliminated through the equation of motion (\ref{eq:EoM-m.}).
In order to be explicit, let us define the equation-of-state parameter
and the adiabatic sound speed,
\begin{eqnarray}
w & \equiv & \frac{\mathcal{P}}{\mathcal{E}}\,,\label{eq:w_ca}\\
c_{\text{a}}^{2} & \equiv & \frac{\dot{\mathcal{P}}}{\dot{\mathcal{E}}}=w-\frac{w'}{3(1+w)}\,,\nonumber 
\end{eqnarray}
where the prime $'$ denotes a differentiation with respect to the
logarithm of the scale factor, $\ln a$ and all the quantities above
are background quantities.%
\footnote{Note that the relationship between $c_{\text{a}}^{2}$ and $w$ is
strictly speaking only valid when no external matter is present: in
the presence of external matter, the DE energy momentum tensor defined
here is not conserved (see eq. (\ref{eq:BDKnoncons})) and therefore
$\dot{\mathcal{E}}$ has a new contribution dependent on $\rho_{\text{ext}}$.
This modifies the expression for $c_{\text{a}}^{2}$.%
} The $w$ of eq.~(\ref{eq:w_ca}) is the equation of state for the
DE; however, in the DM-less scenario being discussed here it is also
the total $w$ of the cosmological expansion. 

We will focus our study on configurations where
\begin{eqnarray}
|\varkappa'| & \ll & 1\,,\label{eq:hier}\\
\beta & \ll & |\varkappa'|\,.\nonumber 
\end{eqnarray}
The first condition implies that the Planck mass be evolving slowly.
The second implies that the external matter, which is universally
coupled in this model, be coupled relatively weakly, so as to avoid
strong violations of the equivalence principle. It also means that
we are not in the well-studied $f(R)$-like regime discussed on page
\pageref{eq:f(R)props}. We are refraining from discussing issues
such as Solar-System tests in this paper, only concentrating on the
formalism, but the requirements (\ref{eq:hier}) should make it more
likely that the parameter space being discussed is not already ruled
out \cite{Pettorino:2012ts}.

We are going to treat the expressions (\ref{eq:E-sec3B}) and (\ref{eq:P-sec3B})
as functions of three independent variables: $\phi$, $m$ and the
expansion $\theta$. In general one would also have to include the
perturbations of the internal degrees of freedom of the external matter.
The Einstein equations provide some constraints between these variables.
In order to obtain the perturbations for the hydrodynamic variables
we are going to vary with respect to all these quantities. Eliminating
the derivative w.r.t. $\phi$ in favour of the time derivative, we
obtain 
\begin{align}
\delta\mathcal{E}= & \dot{\mathcal{E}}\frac{a^{2}\Theta}{k^{2}}-\dot{\varkappa}\left(\delta\theta-3\dot{H}\frac{a^{2}\Theta}{k^{2}}\right)+m\mathcal{E}_{m}A\frac{a^{2}}{k^{2}}\,,\label{eq:dP1-scales}\\
\delta\mathcal{P}= & \dot{\mathcal{P}}\frac{a^{2}\Theta}{k^{2}}-\dot{\varkappa}\left(c_{\text{s}}^{2}-\frac{2}{3}\right)\left(\delta\theta-3\dot{H}\frac{a^{2}\Theta}{k^{2}}\right)+m\mathcal{P}_{m}A\frac{a^{2}}{k^{2}}\,,\nonumber 
\end{align}
where $A$ is the divergence of the acceleration defined in eq. (\ref{eq:A-def}).
It expresses the spatial perturbation of the scalar as seen in the
scalar frame (\ref{eq:u_def}), orthogonal to the perturbation of
the proper time of that frame represented by $\Theta$. We can now
use (\ref{eq:ThetaPert}) to re-express the $\delta\theta$ in terms
of $\Theta$ and $A$. Finally, we have to rewrite every $\Theta$
in terms of the momentum flux $\Xi$, for which the definition eq.
(\ref{eq:XiDef}) reduces in our model (\ref{eq:action1}) to
\begin{equation}
\Xi=(\mathcal{E}+\mathcal{P})\Theta+\dot{\varkappa}A\,.\label{eq:Xi_BDK}
\end{equation}
We do this since we wish to obtain evolution equations purely in terms
of $\delta\mathcal{E}$ and we are able to rewrite the EMT conservation
equation to retain only the combination $\Xi$, eq.~(\ref{eq:EnCons_Xi}).
Putting all these together, and provided $\mathcal{E}+\mathcal{P}\neq0$,
we obtain
\begin{eqnarray}
\delta\mathcal{E} & = & e^{-\varkappa}m^{2}D\left(1+\frac{2\beta}{\mathcal{E}+\mathcal{P}}\frac{k^{2}}{a^{2}}\right)\frac{a^{2}}{k^{2}}A-3\left(\frac{aH}{k}\right)^{2}\frac{\Xi}{H}-\frac{\dot{\varkappa}}{\mathcal{E}+\mathcal{P}}\Xi\,,\label{eq:dEdP-AXi}\\
\delta\mathcal{P} & = & e^{-\varkappa}m^{2}D\left(1+\frac{2\beta}{\mathcal{E}+\mathcal{P}}\frac{k^{2}}{a^{2}}\right)C^{2}\frac{a^{2}}{k^{2}}A-3c_{\text{a}}^{2}\left(\frac{aH}{k}\right)^{2}\frac{\Xi}{H}-\frac{\dot{\varkappa}}{\mathcal{E}+\mathcal{P}}\left(c_{\text{s}}^{2}-\frac{2}{3}\right)\Xi\nonumber 
\end{eqnarray}
where in the coefficients we have neglected a number of terms subdominant
at scales below the horizon. We will return to the definition of the
quantity $C^{2}$ in eq.~(\ref{eq:C2-exact}).\textbf{ }Since we
have no evolution equation for $A$, we have to eliminate the dependence
of $\delta\mathcal{P}$ on it by combining with the equation for $\delta\mathcal{E}$,
which gives us
\begin{alignat}{1}
\delta\mathcal{P} & =C^{2}\delta\mathcal{E}-3(c_{\text{a}}^{2}-C^{2})\left(\frac{aH}{k}\right)^{2}\frac{\Xi}{H}-\frac{\dot{\varkappa}}{\mathcal{E}+\mathcal{P}}\left(c_{\text{s}}^{2}-\frac{2}{3}-C^{2}\right)\Xi+\label{eq:dP-closure}\\
 & +\frac{3}{2}\dot{\varkappa}e^{-\varkappa}\rho\left(c_{\text{s}}^{2}-\frac{2}{3}-C^{2}\right)\frac{a^{2}}{k^{2}}\Theta_{\text{m}}+\beta e^{-\varkappa}\delta\rho\,.\nonumber 
\end{alignat}
where we have for the moment restored the dependence on external matter.
As we have previously claimed in section \ref{sub:GenPerts}, this
is indeed a closure relation which has the structure of eqs (\ref{eq:ClosureParam}).
All the parameters are purely functions of the background, apart from
$C^{2}$, which contains one scale as we shall see below. Nonetheless,
this is a significantly more complex expression than that obtained
for k-\emph{essence }in eq.~(\ref{eq:k-ess-closure-Xi}).

We can perform the same computation for the anisotropic-stress closure
relation. In the model (\ref{eq:action1}), the expression for the
anisotropic stress is simple
\begin{equation}
\frac{k^{2}}{a^{2}}\delta\pi=-\dot{\varkappa}\Theta=\frac{m^{2}D}{1+f}\frac{2\beta}{\mathcal{E}+\mathcal{P}}A-\frac{\dot{\varkappa}}{\mathcal{E}+\mathcal{P}}\Xi\,.\label{eq:piwithA}
\end{equation}
We can again obtain the closure relation of the correct structure
by eliminating the $A$ term from eq.~(\ref{eq:piwithA})
\begin{equation}
\frac{k^{2}}{a^{2}}\delta\pi=\Pi\delta\mathcal{E}+3\Pi\left(\frac{aH}{k}\right)^{2}\frac{\Xi}{H}+(\Pi-1)\frac{\dot{\varkappa}}{\mathcal{E}+\mathcal{P}}\Xi\,,\label{eq:Pi-Closure}
\end{equation}
with
\begin{equation}
\Pi\equiv\frac{2\beta}{\mathcal{E}+\mathcal{P}}\frac{k^{2}}{a^{2}}\left[1+\frac{2\beta}{\mathcal{E}+\mathcal{P}}\frac{k^{2}}{a^{2}}\right]^{-1}\,,\label{eq:PI-def}
\end{equation}
$\Pi$ interpolates between a small $k$-dependent quantity at large
scales (effectively 0, anisotropic stress vanishes) and 1 at small
scales, which signifies that the anisotropic stress be large.\\

Let us return to the final undefined variable in (\ref{eq:dP-closure}),
\begin{equation}
C^{2}=\left[\frac{e^{\varkappa}(m\mathcal{P}_{m}+3Hc_{\text{a}}^{2}\dot{\varkappa})}{m^{2}D}+\frac{2\beta}{\mathcal{E}+\mathcal{P}}\left(c_{\text{s}}^{2}-\frac{2}{3}\right)\frac{k^{2}}{a^{2}}\right]\left[1+\frac{2\beta}{\mathcal{E}+\mathcal{P}}\frac{k^{2}}{a^{2}}\right]^{-1}\,.\label{eq:C2-exact}
\end{equation}
where we have again neglected some terms irrelevant on subhorizon
scales. This expression is clearly not exactly the physical sound
speed of the scalar as was defined through the acoustic metric in
eq. (\ref{eq:cs2}). However, given the assumptions (\ref{eq:hier}),
we can approximate the expression (\ref{eq:C2-exact}) as
\begin{eqnarray}
C^{2} & \simeq & \left[c_{\text{s}}^{2}+M_{C}^{2}+\left(c_{\text{s}}^{2}-\frac{2}{3}\right)\frac{2\beta k^{2}}{a^{2}(\mathcal{E}+\mathcal{P})}\right]\left[1+\frac{2\beta}{\mathcal{E}+\mathcal{P}}\frac{k^{2}}{a^{2}}\right]^{-1}+\mathcal{O}\left(\beta^{2},\beta c_{\text{s}}^{2},\right)\,,\label{eq:C^2}\\
M_{C}^{2} & \equiv & -\frac{2\beta}{\varkappa'}(1-3c_{\text{a}}^{2})+4\beta\left(\frac{\varkappa_{\phi\phi}}{\varkappa_{\phi}^{2}}\right)+\frac{2\beta}{\varkappa'}\frac{E_{\phi}}{mHD}\left(\frac{mD_{m}}{D}-\frac{mE_{\phi m}}{E_{\phi}}\right).\label{eq:MC^2}
\end{eqnarray}
We should stress that the above is an approximation only valid when
the conditions (\ref{eq:hier}) are satisfied and\textbf{ }only on
subhorizon scales.%
\footnote{In particular, these expressions are \emph{not} valid in the limit
of $f(R)$ gravity. One has to go back to the expression for the pressure
(\ref{eq:P-sec3B}) and recalculate assuming conditions (\ref{eq:f(R)props})
to be satisfied.%
} We can see that as the scale $k$ is varied, the quantity $C^{2}$
interpolates between two limiting values, both related to the physical
sound speed, $c_{\text{s}}^{2}$ and $c_{\text{s}}^{2}-2/3$. In addition,
there is a modification resulting from the $M_{C}^{2}$ term which
we should stress is defined here to be dimensionless, effectively
in terms of the Hubble scale. 

As we will see, $M_{C}^{2}$ determines one of the scales in the problem,
which is equivalent to the Compton scale in $f(R)$ models. It represents
the degree of breaking of shift symmetry in the Lagrangian (\ref{eq:action1}).
If we make the $\phi$ dependence of $K(\phi,X)$ and $\varkappa_{\phi}$
sufficiently weak, then only the first term of eq. (\ref{eq:MC^2})
will contribute and $M_{C}^{2}\sim\mathcal{O}(\beta/\varkappa')$.
We will refer to the scale created by $M_{C}^{2}$ in our perturbation
as the Compton scale or wavelength. Interestingly, it appears from
the definition (\ref{eq:MC^2}) that $M_{C}^{2}$ can be of either
sign and therefore that solutions with increasing and decreasing Planck
mass are not equivalent. One of these signs will signify a fast tachyonic
instability for the perturbations implying that the background solution
chosen is inappropriate. We will assume that if the effect of this
mass is at all relevant then the background solution must be such
that the mass is positive. 

In both $C^{2}$ and $\Pi$ there appears the same scale-dependence
which changes the coefficients in the evolution equations for the
energy-density perturbations and therefore qualitatively changes their
evolution. We shall discuss this scale in detail in the following
sections.

\subsubsection{New Transition Scale}

Both the formula for $C^{2}$, eq.~(\ref{eq:C^2}), and for $\Pi$,
eq.~(\ref{eq:PI-def}), feature the same new \emph{transition scale},
\begin{equation}
\left(\frac{k_{\text{T}}}{aH}\right)^{2}\equiv\frac{\mathcal{E}+\mathcal{P}}{2\beta H^{2}}\,=\frac{3(1+w)}{2\beta}.\label{eq:kT}
\end{equation}
As we go across this scale, the coefficients interpolate between two
values and the fluid transitions between two behaviours. eq.\foreignlanguage{english}{~(\ref{eq:PI-def})}
implies that at large scales, $k\ll k_{\text{T}}$, the anisotropic
stress in the fluid will be absent, while at scales $k\gg k_{\text{T}}$
we will have $\Pi\simeq1$ and have anisotropic stress comparable
to the energy density perturbation. Thus the scale $k_{\text{T}}$
delineates the transition between what we will call the \emph{perfect}
and \emph{imperfect }regimes. In the model (\ref{eq:action1}) in
the imperfect regime, the dark-energy fluid \emph{always} carries
anisotropic stress.

This transition scale is determined by the relative size of the k-\emph{essence}
terms as compared to the second-derivative terms in the EMT for linear
perturbations. The k-\emph{essence }terms contain $\delta\phi$ and
$\delta m$, with coefficients related to the function $K(\phi,X)$.
However, the second-derivative terms introduce terms containing $(k/a)^{2}\delta\phi$
and $(k/a)^{2}\delta m$ with a common coefficient $\dot{\varkappa}$.
The scale at which the two sets of terms are comparable defines (\ref{eq:kT}).
Equivalently, the transition between the two regimes is defined by
the scale at which the $\Theta$ and $\delta q$ terms are comparable
in $\Xi$,\textbf{ }eq.~(\ref{eq:XiDef}). Thus in the imperfect
regime, it is the energy flow $q$ terms that determine the evolution
of the scalar-fluid perturbations. It is also worth stressing that
it is the combination $E+P$ that matters here, i.e. only the kinetic
energy stored in the k-\emph{essence} terms, not the potential. This
means, for example, that $f(R)$ models which do not have a kinetic-energy
term in $K(\phi)$ are always in the imperfect regime.\textbf{}\\

It is interesting to note that this scale is only visible through
the EMT and does not appear in the equation of motion (\ref{eq:Scalar_Eom}).
The background evolution of the value of the scalar field $\phi$
is determined by the background equation of motion, which has no spatial
dependence. The evolution of the scalar's perturbations is then determined
by the perturbed equation of motion, which can be written as
\begin{equation}
\ddot{\delta\phi}+3H(1+\gamma)\dot{\delta\phi}+c_{\text{s}}^{2}\frac{k^{2}}{a^{2}}\delta\phi+M^{2}\delta\phi=\dot{\varkappa}\delta\rho\,,\label{eq:PertEOM}
\end{equation}
where we have neglected the gravitational potentials subdominant subhorizon
and have hidden the complexity of the expression in the coefficients
$M^{2}$ and $\gamma$, the formulae for which we won't need. The
gradient term for the scalar is clearly visible here and is determined
by the physical sound speed $c_{\text{s}}^{2}$. Provided a slow-enough
variation of the coefficients, the solutions to these equations are
generically oscillatory. There is a change in behaviour as one crosses
the Jeans scale, but this is the only transition. When external matter
is present, it provides a source to eq.~(\ref{eq:PertEOM}), which
will induce an additional particular solution. This may or may not
be dominant over the homogeneous solutions discussed, depending on
the behaviour of all the coefficients. 

However, in order to understand how the evolution of the scalar-field
perturbation translates into a gravitational effect which would be
felt by external matter one must consider the relation of the scalar
perturbations to the perturbed EMT (\ref{eq:E-sec3B}). And it is
here that a single solution of the equation of motion provides two
distinct behaviours. The\emph{ same} evolution of $\phi$ and $\delta\phi$
in time gives one behaviour to the k-\emph{essence }part of the perturbed
energy density
\[
\delta\mathcal{E}_{\text{k-essence}}\simeq e^{-\varkappa}\left(E_{\phi}\delta\phi+E_{m}\dot{\delta\phi}\right)\,,
\]
and a \emph{different }one to the contribution from the second derivatives,
\[
\delta\mathcal{E}{}_{\text{2-diff}}\simeq-\varkappa_{\phi}\frac{k^{2}}{a^{2}}\delta\phi\,.
\]
Thus the evolution of the energy-density perturbation, and therefore
also the impact on the metric perturbations, can be completely different
above and below the scale $k_{\text{T}}$, despite the same underlying
solution for $\delta\phi$ and $\phi$. 

In the fluid language, we see this transition of behaviour exactly
as a result of the change of the closure relations as we go across
the scale $k_{\text{T}}$. The conservation equations for the EMT
must in the end reduce to the equation of motion for the scalar, whatever
the scale we are looking at. Since the energy-density perturbation
is a different function of $\delta\phi$ at different scales, the
closure relations must adjust the fluid equations in such a way that
the underlying solution for the scalar perturbation is always the
same whatever the scale. Conversely, the fluid equations must predict
a different evolution for the energy-density perturbation at different
scales, given that $\delta\phi$ does not change its behaviour. This
is achieved by changing the closure relations and the coefficients
in the evolution equation for the energy-density perturbation (\ref{eq:dEEvol}),
for example by introducing an anisotropic stress.\\

Let us return to the discussion of the closure relations. Given the
transition scale (\ref{eq:kT}), we can approximate $C^{2}$ either
side of it: 
\begin{alignat}{2}
 & C^{2}\simeq c_{\text{s}}^{2}+M_{C}^{2}\quad &  & \text{for}\, k\ll k_{\text{T}}\,,\label{eq:C2-perf}\\
 & C^{2}\simeq c_{\text{s}}^{2}-\frac{2}{3}+M_{C}^{2}\frac{k_{\text{T}}^{2}}{k^{2}}\quad &  & \text{for}\, k\gg k_{\text{T}}\,.\label{eq:C2-impf}
\end{alignat}
where we have been agnostic as to the size of the Compton term, $M_{C}^{2}$.
According to our parameterisation, the Jeans scale is determined by
$C^{2}+\nicefrac{2}{3}\Pi$, eq. (\ref{eq:dEEvol}). Let us preview
the implications of these approximations, leaving the detailed discussion
for sections \ref{sub:perf} and \ref{sub:impf}:
\begin{itemize}
\item If $c_{\text{s}}^{2}\gg M_{C}^{2}$, the Jeans scale lies inside the
Compton scale. We can neglect $M_{C}^{2}$ which is irrelevant to
the dynamics of perturbations, given an appropriate background solution.
The Jeans term is always determined by the sound speed.
\item If $c_{\text{s}}^{2}\ll M_{C}^{2}$, the Compton scale lies inside
the Jeans scale. eq. (\ref{eq:C2-perf}) implies that in the perfect
regime the perturbation will behave as k-\emph{essence} but with an
effective sound speed equal to $M_{C}^{2}$. The Compton mass term,
as opposed to the sound speed, then determines the frequency of these
oscillations. As we enter the imperfect regime, the Compton mass term
will transition from playing the role of the Jeans term to that of
a usual mass term, continuing to suppress the perturbations. There
will now be a new scale determined by the relative size of $c_{\text{s}}^{2}$
and $M_{C}^{2}$. The physical sound speed will only drive the dynamics
below that scale. This is effectively the picture we have in $f(R)$.
\end{itemize}
We have now obtained both the closure relations, eqs (\ref{eq:dP-closure})
and (\ref{eq:Pi-Closure}) in the necessary form to use in the evolution
equation for energy-density perturbations (\ref{eq:dEEvol}) we have
set up. The forms of parameters $C^{2}$ and $\Pi$ show that there
are two regimes we need to consider, with the transition scale, $k_{\text{T}}$. 

From the closure relations (\ref{eq:dP-closure}) and (\ref{eq:Pi-Closure})
we can now read off the values of the parameters defined in the general
closure relations (\ref{eq:ClosureParam}) for the case of the model
(\ref{eq:action1}). We have presented their values in the two regimes
in Table \ref{tab:ClosureParams}. As can be seen, within each regime,
all the parameters are indeed functions of the background quantities
only, i.e. at most functions of time. Given this reduced set of parameters,
the equivalent of eq. (\ref{eq:dEEvol}) for the model (\ref{eq:action1})
is %
\footnote{We have neglected the $\varpi_{2}$ terms, which we can only do when
$|\varkappa'|\ll|1+w|$, which is implied by our choice (\ref{eq:hier}).
This is not the case for $f(R)$ models (see eq. (\ref{eq:deepMDE})).%
}
\begin{alignat}{1}
\ddot{\delta\mathcal{E}} & +H(8+3c_{\text{a}}^{2}-2\varpi_{1})\dot{\delta\mathcal{E}}+\left(C^{2}+\frac{2}{3}\Pi\right)\frac{k^{2}}{a^{2}}\delta\mathcal{E}+\label{eq:dEEvolBDK}\\
 & +H^{2}\left(4(1-\Pi)\frac{H'}{H}+(3-2\Pi)(5+3c_{\text{a}}^{2})\right)\delta\mathcal{E}=0\,,\nonumber 
\end{alignat}
 Of particular note is the presence of $\varpi_{1}$ in the friction
term. This term interpolates between 0 and 1 and therefore reduces
the friction term on scales where the anisotropic stress is present.
As we show, this leads to slower-decaying homogeneous modes in this
regime.

\begin{table}[th]
\begin{centering}
\begin{tabular}{rrr}
\toprule 
\textsf{\textbf{\emph{Parameter}}} & \textsf{\textbf{\emph{Perfect Regime}}} & \textsf{\textbf{\emph{Imperfect Regime}}}\tabularnewline
\midrule
\midrule 
$C^{2}$ & $c_{\text{s}}^{2}+M_{C}^{2}$ & $c_{\text{s}}^{2}-\frac{2}{3}+\left(M_{C}k_{\text{T}}/k\right)^{2}$\tabularnewline
\midrule 
\textsf{\textbf{$\Pi$}} & 0 & 1\tabularnewline
\midrule 
\textsf{\textbf{$C^{2}+\nicefrac{2}{3}\Pi$}} & $c_{\text{s}}^{2}+M_{C}^{2}$ & $c_{\text{s}}^{2}+\left(M_{C}k_{\text{T}}/k\right)^{2}$\tabularnewline
\midrule 
\textsf{\textbf{$\Sigma_{1}$}} & $C^{2}-c_{\text{a}}^{2}$ & $C^{2}-c_{\text{a}}^{2}$\tabularnewline
\midrule 
\textsf{\textbf{$\varpi_{1}$}} & 0 & 1\tabularnewline
\midrule 
$\Sigma_{2}$ & $-\frac{\varkappa'}{(1+w)}\left(c_{\text{s}}^{2}-C^{2}-\frac{2}{3}\right)$ & 0\tabularnewline
\midrule 
$\varpi_{2}$ & $-\frac{\varkappa'}{(1+w)}$ & 0\tabularnewline
\midrule 
$\beta_{\mathcal{P}}$ & $\beta$ & $\beta$\tabularnewline
\midrule 
$\beta_{\pi}$ & 0 & 0\tabularnewline
\midrule 
$\gamma_{\mathcal{P}}$ & $\frac{3}{2}\varkappa'\left(c_{\text{s}}^{2}-C^{2}-\frac{2}{3}\right)$ & 0\tabularnewline
\midrule 
$\gamma_{\pi}$ & 0 & $-\frac{3}{2}\varkappa'$\tabularnewline
\bottomrule
\end{tabular}
\par\end{centering}

\caption{\label{tab:ClosureParams}Closure parameters for the model (\ref{eq:action1})
as obtained by comparing the closure relations (\ref{eq:dP-closure})
and (\ref{eq:Pi-Closure}) with the general structure proposed in
eqs (\ref{eq:ClosureParam}). The model contains two regimes, perfect
and imperfect, separated by the transition scale $k_{\text{T}}$,
eq.~(\ref{eq:kT}). In the expressions for $C^{2}$, we have assumed
that the external matter constitutes a negligible part of the total
energy density: i.e. dark-energy dominates. This table is in fact
exact apart from this result for $C^{2}$, where the approximation
(\ref{eq:C^2}) would need to be recalculated when matter is present
or conditions (\ref{eq:hier}) do not hold. In the presence of external
matter one would need to also take into account the non-conservation
of the DE EMT, eqs (\ref{eq:NonConsPert}). We have provided the results
for the coefficients $\beta_{i}$ and $\gamma_{i}$ for completeness.}
\end{table}

\subsubsection[Perfect regime, $k\ll k_{\text{T}}$]{Perfect regime, $\boldsymbol{k}\boldsymbol{\ll}\boldsymbol{k}_{\text{T}}$\label{sub:perf}}

In this regime, the closure relations (\ref{eq:dP-closure}) and (\ref{eq:piwithA})
reduce to
\begin{eqnarray}
\delta\mathcal{P} & = & C^{2}\delta\mathcal{E}+3(C^{2}-c_{\text{a}}^{2})\frac{\Xi}{H}\left(\frac{aH}{k}\right)^{2},\label{eq:perf_close}\\
\frac{k^{2}}{a^{2}}\delta\pi & \simeq & 0\,,\nonumber \\
C^{2} & = & c_{\text{s}}^{2}+M_{C}^{2}.
\end{eqnarray}
These closure relations take the same form that generalised k-\emph{essence}
does (with the important replacement of $c_{\text{s}}^{2}$ with $C^{2}$),
see eq.~(\ref{eq:k-ess-closure-Xi}). The evolution equation for
energy-density perturbations (\ref{eq:dEEvolBDK}) reduces to
\begin{alignat}{1}
\ddot{\delta\mathcal{E}} & +H(8+3c_{\text{a}}^{2})\dot{\delta\mathcal{E}}+C^{2}\frac{k^{2}}{a^{2}}\delta\mathcal{E}+3H^{2}\left[5+3c_{\text{a}}^{2}-2(1+w)\right]\delta\mathcal{E}=0\,,\label{eq:dE-evol-perf}
\end{alignat}
The only correction to the standard k-\emph{essence} behaviour is
the fact that if the Compton wavelength is inside the Jeans length,
$M_{C}^{2}>c_{\text{s}}^{2}$, it will determine the behaviour of
the perturbations, determining the oscillation frequency of the solutions.
We can rewrite the above as an equation for the evolution of the density
contrast, 
\begin{alignat}{1}
\delta'' & +\left(\frac{1}{2}-\frac{9}{2}w-\frac{w'}{1+w}\right)\delta'+C^{2}\frac{k^{2}}{a^{2}H^{2}}\delta+M^{2}\delta=0\,,\label{eq:d-k-subdom-1}\\
M^{2} & =-\frac{3}{2}\left(1+2w-3w^{2}+\frac{2w'}{1+w}\right)\,,\nonumber 
\end{alignat}
with $\delta\equiv\delta\mathcal{E}/\mathcal{E}$. We can now obtain
the standard solutions to the homogeneous equations in the approximation
that $w$ is slowly varying. The Jeans length in the perfect regime
is determined by the larger of $c_{\text{s}}^{2}$ and $M_{C}^{2}$,
\[
\left(\frac{k_{\text{Jeans}}}{aH}\right)^{2}=\frac{\left|M^{2}\right|}{C^{2}}.
\]

\begin{enumerate}
\item \emph{Above Jeans length, }$k\ll k_{\text{Jeans}}$. The tachyonic
mass $M^{2}$ dominates the mass term allowing the DE perturbations
to evolve as a power law, with modes 
\begin{equation}
\delta_{+}\propto a^{1+3w}\qquad\text{and}\qquad\delta_{-}\propto a^{-\frac{3}{2}(1-w)}\,.\label{eq:k-dom-modes-1}
\end{equation}
We therefore have a growing mode only for $w>-1/3$. For $w>1$, both
the modes are growing. In this regime, during DE domination, the fluid
will only cluster when the universe is not accelerating. The gravitational
potential is constant.
\item \emph{Below Jeans length}, $k\gg k_{\text{Jeans}}$. At scales smaller
than the Jeans length $Ck/aH\gg1$, pressure support will arrest any
potential collapse, leading to oscillating solutions,
\begin{eqnarray*}
\delta & = & a^{-(1-9w)/4}\left(A_{1}J_{n}\left(\frac{2Ck}{(1+3w)aH}\right)+A_{2}J_{-n}\left(\frac{2Ck}{(1+3w)aH}\right)\right)\,,\\
 &  & n\equiv\frac{1-9w}{2(1+3w)}\,.
\end{eqnarray*}
with $A_{1,2}$ constants of integration and $J_{n}$ Bessel functions
of the first kind. Note that the horizon is shrinking during acceleration,
so the modes will eventually leave this regime and begin to behave
as if outside the Jeans length. Depending on the hierarchy between
$c_{\text{s}}^{2}$ and $M_{C}^{2}$, the oscillation frequency will
be determined by the Compton term or the sound speed.
\end{enumerate}
These results match those studies in refs \cite{Christopherson:2008ry,Creminelli:2008wc,Creminelli:2009mu,Sapone:2009mb}
for the case of k-\emph{essence.}

\subsubsection[Imperfect regime, $k\gg k_{\text{T}}$]{Imperfect regime, $\boldsymbol{k}\boldsymbol{\gg}\boldsymbol{k}_{\text{T}}$\label{sub:impf}}

In this regime in the absence of matter, the closure relations (\ref{eq:dP-closure})
and (\ref{eq:piwithA}) reduce to
\begin{eqnarray}
\delta\mathcal{P} & = & \left(c_{\text{s}}^{2}-\frac{2}{3}+M_{C}^{2}\left(\frac{k_{\text{T}}}{k}\right)^{2}\right)\delta\mathcal{E}+3\left(c_{\text{s}}^{2}-\frac{2}{3}-c_{\text{a}}^{2}+M_{C}^{2}\left(\frac{k_{\text{T}}}{k}\right)^{2}\right)\frac{\Xi}{H}\left(\frac{aH}{k}\right)^{2},\label{eq:dP-closure-impf}\\
\frac{k^{2}}{a^{2}}\delta\pi & = & \delta\mathcal{E}+3\frac{\Xi}{H}\left(\frac{aH}{k}\right)^{2}\,.\label{eq:dpi-closure-impf}
\end{eqnarray}
It is important to stress that the closure relations (\ref{eq:dP-closure-impf})
are \emph{not} hydrodynamic. Firstly, the mass terms $M_{C}^{2}$
appear in eq.~(\ref{eq:dP-closure-impf}). If the Compton wavelength
lies outside of the Jeans length, then the mass terms are always irrelevant
in this regime. However, if the opposite is true, then the mass terms
will dominate the dynamics on the outer edge of this regime.

Secondly, it is very striking that, at small scales, where one would
naively expect to recover the sound speed, the ratio $\delta\mathcal{P}/\delta\mathcal{E}$
can be negative when the physical sound speed $c_{\text{s}}^{2}$
is small. If one were to take this as a definition of the speed of
sound, one would interpret this sort of configuration as unstable.
However, as we have been showing, the EMT of a non-minimally-coupled
scalar field features anisotropic stress. This stress is very large,
since in the imperfect regime it is of the same order as the energy-density
perturbation, and it acts in such a way so as to precisely cancel
out the term which would contribute to the would-be instability.%
\footnote{The appropriate size comparison is between dimension-four quantities
$\delta\mathcal{E}$ and $\frac{k^{2}}{a^{2}}\delta\pi$, as in eq.
(\ref{eq:dpi-closure-impf}).%
} 

This anisotropic stress from the perspective of hydrodynamics represents
shear viscosity. This is a term appearing as a first-order correction
in the gradient expansion and one would expect it to be small, lest
it signify that the gradient expansion itself were invalid. Here,
there are no terms higher-order in the gradient expansion to worry
about: the elements of the EMT are at most exactly second order in
gradients. These closure relations are a full rearrangement of the
equation of motion for the scalar field at linear order in perturbations.

It is actually not so surprising that it is the combination of pressure
and anisotropy perturbations that provides the support against collapse.
Projecting out the spatial trace of the perturbed Einstein equations
(the $G_{ii}$ components, effectively), we obtain 
\[
2\ddot{\Phi}+6H\dot{\Phi}-2H\dot{\Psi}-(6H^{2}+4\dot{H})\Psi=-\left(\delta\mathcal{P}+\frac{2}{3}\frac{k^{2}}{a^{2}}\delta\pi\right)\,,
\]
after substituting for a scale-dependent term in $G_{ii}$ with the
anisotropic stress. Thus the time evolution of the potentials reacts
to the whole term in the parentheses, both pressure and anisotropic
stress.

We should also note that the relationship between anisotropy and the
energy perturbation of eq. (\ref{eq:dP-closure-impf}) is the standard
result for scalar-tensor theories: the nature of the anisotropic stress
generated by the non-minimal coupling to gravity is such that it cancels
the second-derivative contribution to the energy density leaving just
the k\emph{-essence }part of the energy-density perturbation. Effectively,
in the imperfect regime, the dark-energy\emph{ }density perturbations
in this model generate no lensing potential:
\begin{equation}
\frac{k^{2}}{a^{2}}(\Phi-\Psi)=\delta\mathcal{E}-\frac{k^{2}}{a^{2}}\delta\pi\simeq\delta\mathcal{E_{\text{k-essence}}}\ll\frac{k^{2}}{a^{2}}\Phi\,.\label{eq:lensing}
\end{equation}
This property is an intrinsic property of this DE fluid and is always
valid inside the imperfect regime, whatever the configuration. It
is not just limited to quasi-static solutions. Some more general scalar-field
theories have a coupling to gravity that cannot be conformally rescaled
away and therefore the photons are sensitive to the energy perturbations
of the scalar field (say the Lagrangians $\mathcal{L}_{4}$ and $\mathcal{L}_{5}$
in ref.~\cite{DeFelice:2011hq}).

The evolution equation (\ref{eq:dEEvolBDK}) in the imperfect regime
reduces to 
\begin{alignat}{1}
\ddot{\delta\mathcal{E}} & +3H(2+c_{\text{a}}^{2})\dot{\delta\mathcal{E}}+c_{\text{s}}^{2}\frac{k^{2}}{a^{2}}\delta\mathcal{E}+H^{2}\left[\frac{3}{2}(1+w)\frac{M_{C}^{2}}{\beta}+5+3c_{\text{a}}^{2}\right]\delta\mathcal{E}=0\,,\label{eq:de-Evol-impf}
\end{alignat}
We should stress that the coefficient of the Jeans term here is just
the physical sound speed $c_{\text{s}}^{2}$. There are two significant
differences between this equation and its equivalent in the perfect
regime (\ref{eq:dE-evol-perf}): firstly the coefficient of the friction
term is reduced from $H(8+3c_{\text{a}}^{2})$ to $H(6+3c_{\text{a}}^{2})$
as a result of the presence of anisotropic stress. This leads to a
significant change in the evolution of the solutions of the homogeneous
modes with the scale factor and can affect the relative importance
of the solutions to the homogeneous equation and the particular solutions
in the presence of external matter. Secondly, in the imperfect regime,
the effective mass of the mode has contributions from the Jeans term,
the Compton term and the other terms depending only on the background
expansion, we will therefore have three regimes for perturbation evolution
in this case. 

We can now rewrite eq. (\ref{eq:de-Evol-impf}) as an equation for
the density contrast:
\begin{alignat}{1}
\delta'' & -\frac{3}{2}\left(1+3w-\frac{2w'}{3(1+w)}\right)\delta'+c_{\text{s}}^{2}\frac{k^{2}}{a^{2}H^{2}}\delta+M^{2}\delta=0\,,\label{eq:DEevol_impf}\\
M^{2} & =\frac{1}{2}\left((1+3w)^{2}+3(1+w)\frac{M_{C}^{2}}{\beta}-\frac{2w'}{1+w}\right).\nonumber 
\end{alignat}
We then have three potential classes of solutions, which depend on
our comparing the relative sizes of the terms arising from
\begin{alignat}{5}
 & \text{Compton} &  & \text{gravitational instability} &  &  &  & \text{\quad Jeans}\nonumber \\
 & \quad M_{C}^{2} & \leftrightarrow & \qquad\frac{\beta(1+3w)^{2}}{1+w} &  & \leftrightarrow &  & \quad c_{\text{s}}^{2}\frac{k^{2}}{k_{\text{T}}^{2}} & \,.\label{eq:MassContrib}
\end{alignat}
The gravitational instability term is the imperfect-regime equivalent
of the tachyonic mass for a perfect fluid. We have kept the standard
name, even though it could be considered inappropriate here since
this mass is always positive, irrespective of $w$. As in the perfect
regime, if $c_{\text{s}}^{2}\gg M_{C}^{2}$ then the Compton term
will have no effect on the perturbations at all. On the other hand,
the dominance of the Compton term over the gravitational instability
or the Jeans terms implies that we are outside the Compton wavelength
of the field and its mass is relevant for the dynamics, shielding
the interactions. At small enough scales, the Jeans term will end
up dominating eventually, since there are always scales shorter than
the Compton wavelength. However whether there exists a region of scales
where the gravitational-instability terms dominate is just a question
of the particular model and the background configuration and not scales.
We should note that it is possible for the physical sound speed to
be large enough for the Jeans length to lie already inside the perfect
regime. In this is the case and $M_{C}^{2}\ll c_{\text{s}}^{2}$,
the imperfect regime will only feature Jeans-term dominated scales.\\

We can now solve for the evolution of the DE perturbations in each
of the three cases: Jeans term, Compton term and gravitational-term
domination.
\begin{enumerate}
\item \emph{Gravitational instability terms dominate: $\beta(1+3w)^{2}/(1+w)\gg M_{C}^{2},c_{\text{s}}^{2}k^{2}/k_{\text{T}}^{2}$.
}The DE perturbations evolve as a power law with
\begin{equation}
\delta_{+}\propto a^{(1+3w)/2},\qquad\delta_{-}\propto a^{(1+3w)}\,.\label{eq:impfmodes}
\end{equation}
Just as in the case of the perfect regime, the homogeneous modes decay
whenever the DE is causing acceleration, $w>-1/3$. However, here
the mode which decays quicker does so only as fast as the ``growing''
mode in the case of the perfect fluid. There is now a new mode which
decays even more slowly than in the case of the perfect fluid, c.f.~eq.~(\ref{eq:k-dom-modes-1}).
This means that generically the Newtonian potential will be growing
in this regime whenever the background is accelerating. The Poisson
equation implies that the Newtonian potential. 
\[
\Phi=\frac{3}{2}\frac{a^{2}H^{2}}{k^{2}}\delta_{+}\propto a^{-(1+3w)/2}\,.
\]
One needs to be careful, since these equations are only valid subhorizon
and inside the imperfect regime, while the modes are leaving both
and our universe is not yet in a state completely dominated by dark
energy. However, this diametrically different behaviour could significantly
affect late-time potentials already today. We will return to study
these questions in detail in a separate work. 
\item \emph{Compton term dominates, $M_{C}^{2}\gg\beta(1+3w)^{2}/(1+w),c_{\text{s}}^{2}k^{2}/k_{\text{T}}^{2}$}
. The behaviour of the modes depends on the overall sign of $(1+w)M_{C}^{2}/\beta$.
Assuming that the Compton term be constant, we then have two solutions,
\begin{alignat*}{2}
\frac{(1+w)}{\beta}M_{C}^{2}>0, & \qquad &  & \delta=a^{\frac{3}{4}(1+3w)}\left(A_{1}\cos\frac{3M_{C}^{2}}{2\beta}\ln a+A_{2}\cos\frac{3M_{C}^{2}}{2\beta}\ln a\right)\,,\\
\frac{(1+w)}{\beta}M_{C}^{2}<0, & \qquad &  & \delta=A_{1}a^{p_{+}}+A_{2}a^{p_{-}}\,,\qquad p_{\pm}\simeq\mp\frac{3}{2}\frac{(1+w)}{\beta}M_{C}^{2}\,,
\end{alignat*}
a decaying oscillator or a growing mode depending on the sign. In
either case the gravitational potential grows for an accelerating
equation of state. If the Compton term is substantial, the tachyonic
solution is a very fast instability signifying that the chosen background
solution itself is unstable and should not be considered a realistic
phenomenology.
\item \emph{Jeans term dominates, $c_{\text{s}}^{2}k^{2}/k_{\text{T}}^{2}\gg M_{C}^{2},\beta(1+3w)^{2}/(1+w)$.}
The solution in this regime is an oscillatory function, settling to
a constant amplitude when $w<-1/3$. The leading-order behaviour at
late times is
\[
\delta=A_{1}\cos\left(\frac{2c_{\text{s}}}{1+3w}\frac{k}{aH}\right)+A_{2}\sin\left(\frac{2c_{\text{s}}}{1+3w}\frac{k}{aH}\right)\,.
\]
Thus contrary to the standard perfect-fluid behaviour, the density
perturbation sub-Jeans-horizon does not decay away. Again, the Newtonian
potential in this regime would grow while oscillating, until the mode
exits the sound horizon and moves into regime (1) or (2).
\end{enumerate}
We have shown that at all the scales inside the imperfect regime,
the perturbations evolve is such a way that the gravitational potential
being driven by them grows. This is of course a frame effect: in the
Einstein frame, where gravity is minimally coupled no such increase
would be observed. However, if we assume that the observable baryons
are minimally coupled in the Jordan frame discussed here, they would
indeed see a deepening potential in this late, dark-energy dominated
era. However, in an accelerating universe, modes are exiting the horizon.
Thus any mode starting off inside the Jeans length in the imperfect
regime, will exit the imperfect regime eventually and would leave
the sound horizon. Thus these sort of solutions should be thought
of as being temporary for each mode, with shorter modes spending longer
inside the imperfect regime and therefore inducing a scale-dependence
in the dark-energy perturbation amplitude.

\section{Discussion and conclusion\label{sec:Conc}}

In this paper we have proposed a unified prescription for studying
the subhorizon evolution of linear perturbations in theories of dark
energy comprising a single scalar, including, for example, modified
gravity models such as $f(R)$ and other models containing second-derivative
terms in their EMTs. These EMTs are in general not of perfect-fluid
form.

k-\emph{essence} is the most general scalar theory that has an EMT
of perfect-fluid form. The evolution of linear cosmological perturbations
in this class of models is already well understood. However, any further
modifications to the scalar's Lagrangian, such as galileon terms or
even a simple non-minimal coupling to gravity, generate terms including
second derivatives of the scalar in the EMT. The presence of the second-order
derivatives makes it impossible to express the EMT in perfect-fluid
form. A standard method for dealing with such models, as carried out
in the literature, is to turn to the equation of motion for the scalar,
usually concentrating on the limit of negligible time derivatives
for the scalar field. We have argued that this is in fact on one hand
too specific. For example, for k\emph{-essence} any two background
solutions with the same history of equation-of-state parameter $w$
and sound speed $c_{\text{s}}^{2}$ will give exactly the same evolution
of linear perturbations, irrespective of the actual Lagrangian, as
eq.~(\ref{eq:d-k-subdom-1}) shows. One may of course prefer one
of the Lagrangians over another on the basis of naturalness, but this
is not an observable if we only have access to the cosmological background
and the effect of linear density perturbations on gravitational potentials.
On the other hand: the neglecting of the time derivatives could prove
dangerous, if the coefficient of the friction term in the density-contrast
evolution equation turns out to be small enough to allow for modes
which grow sufficiently quickly on a particular cosmological background.
We have, for example, explicitly demonstrated that in the non-minimally
coupled k-\emph{essence} model, the DE energy-density contrast during
the DE-domination era evolves in such a way so as to cause the gravitational
potential to grow, eq.~(\ref{eq:impfmodes}). These modes should
dominate the gravitational potential at late-enough times on at least
some of the scales.

Our prescription allows for the study of such general models in terms
of the intuitive fluid language instead of dealing directly with the
equation of motion for the scalar field. Since the fluid variables
such as pressure and energy density are just appropriately reorganised
components of the EMT, they will contain all the information to which
the gravitational field is sensitive.

As is well known, one can obtain the evolution equation for the energy-density
perturbations from the perturbed conservation equations for the EMT.
However, two closure relations need to be provided to do this: we
need to relate both the pressure perturbation and the anisotropic-stress
perturbation to the energy-density perturbation. The main result of
this paper is the method for obtaining these closure relations, based
on considering the actual degrees of freedom on a spatial hypersurface. 

As we have explicitly shown in our example, this method produces closure
relations that are not those of perfect-fluid hydrodynamics. One \emph{cannot
}assume that the relationship between the pressure and energy-density
perturbations is hydrodynamical, $\delta\mathcal{P}=c_{\text{H}}^{2}\delta\mathcal{E}$,
with $c_{\text{H}}$ a hydrodynamical sound speed for pressure waves,
as is frequently done. We have explained in detail in sections \ref{sub:hydro}
and \ref{sub:k-ess} why even general k-\emph{essence} models should
not be thought of as a hydrodynamical fluid. The result is that the
closure relations for general scalar-field theories take the somewhat
complicated form (\ref{eq:ClosureParam}) and cannot be reduced to
a simple relationship. 

One may argue that this hydrodynamical relation between $\delta\mathcal{P}$
and $\delta\mathcal{E}$ is the definition of the sound speed for
pressure waves, $c_{\text{H}}$. However, in general this is a not
a physically meaningful quantity: in the example we have studied,
where the anisotropic stress does not vanish, the Jeans length is
not at all related to the quantity $c_{\text{H}}$, but is determined
by the speed of propagation of perturbations of the scalar field,
$c_{\text{s}}$, and therefore causality. As we explicitly show, $c_{\text{s}}$
is obtained by considering the effective wave operator in the scalar
perturbed equation of motion and is not immediately apparent in the
expressions for the pressure perturbation. Even when $c_{\text{s}}$
is simple and constant, the complexity of the closure relations we
have derived shows that $c_{\text{H}}$ could be an arbitrary function
of time and space and therefore not at all a useful parameter for
the description of the perturbations.\\

In order to aid the calculation of the closure relations, we have
introduced a book-keeping parameterisation for them, suitable for
a wide range of models containing a single scalar degree of freedom.
This has allowed us to derive a general equation for the evolution
of DE density perturbations in terms of these parameters which can
be calculated for any particular class of scalar Lagrangians. For
the purpose of illustration, we have shown explicitly how to calculate
the coefficients of the closure relations (\ref{eq:ClosureParam})
in a toy class of models, presenting the results in table \ref{tab:ClosureParams}.
All the parameters which are non-vanishing for this class of Lagrangians
are expressible in terms of physical quantities: $w$, $c_{\text{s}}^{2}$,
the coupling to external matter $\beta$ and the Compton term $M_{C}^{2}$,
each of which is at most a function of time but never scale. Non-minimally
coupled k-\emph{essence} was chosen as the toy model since it is the
simplest one to contain the majority of the features which are present
in more-general scalar theories such as galileons. This prescription
can be directly applied in the study of those more-complex models.

All models containing terms with second derivatives in the EMT also
contain a new scale $k_{\text{T}}$, eq.~(\ref{eq:kT}), determined
by the relative contribution of the k-\emph{essence }and second-derivative
terms in the perturbations. In our toy example, as we move across
this scale, the DE fluid transitions from k-\emph{essence}-like perfect-fluid
behaviour at large subhorizon scales to an imperfect fluid carrying
anisotropic stress at small scales.\textbf{ }Interestingly, the scale
$k_{\text{T}}$ does \emph{not} appear in the perturbed equation of
motion for the scalar itself, just in the EMT. On both sides of the
transition scale, the scalar-field perturbation $\delta\phi$ evolves
in the same manner, at least provided there isn't a scale dependence
in any sources on the external-matter side. The changes in the pressure
and anisotropy closure relations as we cross the scale $k_{\text{T}}$
reflect simply that given the same evolution in time of $\delta\phi$,
the density perturbation $\delta\mathcal{E}$ must evolve differently,
since the \emph{k-essence }and the second-derivative terms have a
completely different functional dependence on $\delta\phi$. Since
it is $\delta\mathcal{E}$ that sources the gravitational potential,
this transition scale is physical. In addition to the transition scale,
the dark energy will feature the known Jeans scale, controlled by
the speed of propagation of small perturbations of the scalar field
and a Compton scale related to the mass of the field. We have not
studied superhorizon behaviour and the transition towards the horizon,
but since the new phenomenology is related to the presence of second
derivatives in the EMT, it will be irrelevant at large enough scales,
where the standard k-\emph{essence }behaviour is restored. Indeed
sufficiently superhorizon the curvature perturbation is conversed
in the standard way \cite{Gao:2011mz}.

What our discussion shows is that it is more appropriate to think
of the anisotropic stress as being related through the closure relation
(\ref{eq:ClosureParam}) to the density perturbation of the DE fluid
rather than directly to the gravitational potentials. In general,
the gravitational potential will also have contributions from other
fluids present in the universe. The density perturbations in any particular
fluid evolve according to the conservation equations for that fluid,
and therefore in principle are independent dynamical variables. In
our example model, in the imperfect regime, the closure relations
imply that $k^{2}\delta\pi/a^{2}\simeq\delta\mathcal{E}$, \emph{irrespective}
of the configuration of the external matter. It is this property of
the fluid that makes the lensing potential independent of the perturbation
in the scalar field, eq.~(\ref{eq:lensing}). And it is only this
relation that is in general independent of the other constituents
of the universe.

The only reason why in $f(R)$ models this scale $k_{\text{T}}$ is
not seen, is that they are a special subclass of the non-minimally
coupled k\emph{-essence }models\emph{ }where the k-\emph{essence}
terms are absent. Thus the DE fluid is imperfect at all scales. The
Compton scale determines whether for a particular mode the energy-density
perturbations are large or small. At scales inside the Compton radius
they are large and therefore the anisotropic stress is large. Outside
--- they are small and so is the anisotropic stress. 

We should stress that every \emph{class }of Lagrangian terms which
generates second derivatives in the EMT (e.g. kinetic gravity braiding,
etc.) will produce such a transition scale determined by the relative
size of its coefficient in the background solution. Each class of
terms will presumably cause slightly different physical effects and
therefore each of these transitions should be potentially observable.

Finally, we should reiterate that in order to simplify the exposition,
we have mostly neglected in this presentation the effect of external
matter. This has allowed us to concentrate on understanding the properties
of the DE fluid itself. As a result of the non-minimal coupling to
gravity, the DE and dark matter are coupled in this model. As we have
previewed in the closure relations (\ref{eq:ClosureParam}), the coupling
to dark matter causes the DM perturbation to appear in the expression
for $\delta\mathcal{P}$ and therefore in the Jeans term. The DM perturbations
will in this way provide a scale-dependent source in the evolution
equation for the energy-density perturbation $\delta\mathcal{E}$,
dominant in the imperfect regime. This ensures that the density perturbations
for DE are large in the imperfect regime, providing a significant
modification to the gravitational potentials in which dark matter
propagates.

The presence of an external energy density provides an additional
complication: the way we defined our DE EMT means that in this case
it is not conserved whenever the effective Planck mass evolves, i.e.
in models with non-minimal coupling to gravity. As we show in the
expressions for the conservation of the EMT (\ref{eq:EnConsPara})
and (\ref{eq:MomConsPara}), there are non-conservation terms which
depend on the scalar and the dark-matter configuration. Therefore,
we will have to provide a similar closure relation to (\ref{eq:ClosureParam}),
one for each of energy and momentum conservation equations. We will
return to this also feature-rich discussion in a separate work.

\section*{Acknowledgements}

It is a pleasure to thank Thorsten Battefeld, Guillermo Ballesteros,
Tommaso Giannantonio, Mark Hindmarsh, Tomi Koivisto, Valerio Marra,
Mariele Motta, Savvas Nesseris, Nelson Nunes, Nicolas Tessore, Wessel
Valkenburg, Alex Vikman for helpful conversations. I.S.~thanks Deutsche
Bahn AG for their on-board conveniences. L.A., M.K.~and I.S.~thank
the Centro de Sciencias de Benasque for their hospitality during the
final stages of preparation of this manuscript. I.D.S.~would like
to thank the ITP Heidelberg for the warm hospitality during the early
stages of this work. The work of L.A.~and I.S.~is supported by the
DFG through TRR33 ``The Dark Universe''. M.K.~acknowledges funding
by the Swiss National Science Foundation.

\appendix

\section{Summary of Notation}

In order to aid the reader, we have compiled the notation used in
this paper in table \ref{tab:Notation}\textbf{.}

\newpage{}

\begin{table}[H]
\begin{tabular}{llll}
\toprule 
\multicolumn{2}{c}{\textsf{\textbf{\textit{\scriptsize Symbol}}}} &  & \tabularnewline
\cmidrule{1-2} 
\textsf{\textbf{\textit{\scriptsize Cov./Bkgd.}}} & \textsf{\textbf{\textit{\scriptsize Perturbation}}} & \textsf{\textbf{\textit{\scriptsize Description}}} & \textsf{\textbf{\textit{\scriptsize Defined in Eq.}}}\tabularnewline
\midrule
\midrule 
\multicolumn{4}{l}{\textsf{\textbf{\textit{\footnotesize Dark-energy variables}}}}\tabularnewline
\midrule 
\textsf{\textbf{\textit{\footnotesize $\phi$}}} & \textsf{\textbf{\textit{\footnotesize $\delta\phi$}}} & {\footnotesize Scalar field, dark energy} & \tabularnewline
\midrule 
\textsf{\textbf{\textit{\footnotesize $X$}}} &  & {\footnotesize Scalar canonical kinetic term} & \tabularnewline
\midrule 
\textsf{\textbf{\textit{\footnotesize $m=\sqrt{2X}$}}} & \textsf{\textbf{\textit{\footnotesize $\delta m$}}} & {\footnotesize Scalar chemical potential} & {\footnotesize (\ref{eq:def_m})}\tabularnewline
\midrule 
\textsf{\textbf{\textit{\footnotesize $\varkappa(\phi)$}}} & \textsf{\textbf{\textit{\footnotesize $\delta\varkappa$}}} & {\footnotesize Non-minimal coupling to gravity} & {\footnotesize (\ref{eq:Einstein!+f})}\tabularnewline
\midrule 
\textsf{\textbf{\textit{\footnotesize $\beta$}}} &  & {\footnotesize Coupling of DE scalar to external matter} & {\footnotesize (\ref{eq:cs2})}\tabularnewline
\midrule 
\multicolumn{4}{l}{\textsf{\textbf{\textit{\footnotesize External EMT}}}}\tabularnewline
\midrule 
{\footnotesize $\rho_{\text{ext}}$/($\rho$)} & {\footnotesize $\delta\rho_{\text{ext}}$/($\delta\rho$)} & {\footnotesize Energy density of matter external to the DE (dark matter)} & {\footnotesize (\ref{eq:extEMT})}\tabularnewline
\midrule 
{\footnotesize $ $$p_{\text{ext}}$} & {\footnotesize $\delta p_{\text{ext}}$} & {\footnotesize Pressure of matter external to the DE} & {\footnotesize (\ref{eq:extEMT})}\tabularnewline
\midrule 
\multicolumn{4}{l}{\textsf{\textbf{\textit{\footnotesize Covariant Kinematical Quantities}}}}\tabularnewline
\midrule 
\textsf{\textbf{\textit{\footnotesize $u^{\mu}$}}} &  & {\footnotesize Velocity of observer measuring EMT} & {\footnotesize (\ref{eq:u_def})}\tabularnewline
\midrule 
\textsf{\textbf{\textit{\footnotesize $\perp_{\mu\nu}$}}} &  & {\footnotesize Spatial metric for observer $u_{\mu}$} & {\footnotesize (\ref{eq:perp_def})}\tabularnewline
\midrule 
\textsf{\textbf{\textit{\footnotesize $\theta$}}} &  & {\footnotesize Expansion } & {\footnotesize (\ref{eq:grad_u})}\tabularnewline
\midrule 
\textsf{\textbf{\textit{\footnotesize $a^{\mu}$}}} &  & {\footnotesize Acceleration of $u^{\mu}$} & {\footnotesize (\ref{eq:accel_def})}\tabularnewline
\midrule 
\textsf{\textbf{\textit{\footnotesize $\sigma_{\mu\nu}$}}} &  & {\footnotesize Shear of $u_{\mu}$} & {\footnotesize (\ref{eq:shear_def})}\tabularnewline
\midrule 
\textsf{\textbf{\textit{\footnotesize $\mathrm{d}/\mathrm{d}\tau$}}} &  & {\footnotesize Derivative w.r.t. to rest-frame time of $u^{\mu}$} & {\footnotesize (\ref{eq:time_deriv_def})}\tabularnewline
\midrule 
\textsf{\textbf{\textit{\footnotesize $(\ )^{\cdot}$}}} &  & {\footnotesize Derivative w.r.t. to coordinate time $t$} & \tabularnewline
\midrule 
\textsf{\textbf{\textit{\footnotesize $\boldsymbol{\overline{\nabla}}_{\lambda}$}}} &  & {\footnotesize Spatial derivative for observer $u^{\mu}$, compatible
with $\perp_{\mu\nu}$} & {\footnotesize (\ref{eq:deriv_decomp_def})}\tabularnewline
\midrule 
\multicolumn{4}{l}{\textsf{\textbf{\textit{\footnotesize Covariant Decomposition of DE
EMT}}}}\tabularnewline
\midrule 
\textsf{\textbf{\textit{\footnotesize $\mathcal{E}$}}} & \textsf{\textbf{\textit{\footnotesize $\delta\mathcal{E}$}}} & {\footnotesize DE energy density} & {\footnotesize (\ref{eq:Tmunu_decomp})}\tabularnewline
\midrule 
\textsf{\textbf{\textit{\footnotesize $E$}}} & \textsf{\textbf{\textit{\footnotesize $\delta E$}}} & {\footnotesize k-}\emph{\footnotesize essence}{\footnotesize{} part
of energy density} & {\footnotesize (\ref{eq:EMT_fluid_decomp})}\tabularnewline
\midrule 
\textsf{\textbf{\textit{\footnotesize $\mathcal{P}$}}} & \textsf{\textbf{\textit{\footnotesize $\delta\mathcal{P}$}}} & {\footnotesize DE pressure} & %
{\footnotesize (\ref{eq:Tmunu_decomp})}%
\tabularnewline
\midrule 
\textsf{\textbf{\textit{\footnotesize $P$}}} & \textsf{\textbf{\textit{\footnotesize $\delta P$}}} & {\footnotesize k-essence part of pressure} & {\footnotesize (\ref{eq:EMT_fluid_decomp})}\tabularnewline
\midrule 
\textsf{\textbf{\textit{\footnotesize $q_{\mu}$}}} &  & {\footnotesize Energy flux vector} & {\footnotesize (\ref{eq:Tmunu_decomp})}\tabularnewline
\midrule 
\textsf{\textbf{\textit{\footnotesize $q$}}} & \textsf{\textbf{\textit{\footnotesize $\delta q$}}} & {\footnotesize Potential for energy flux} & {\footnotesize (\ref{eq:potentials})}\tabularnewline
\midrule 
\textsf{\textbf{\textit{\footnotesize $\tau_{\mu\nu}$}}} &  & {\footnotesize Viscous stress tensor} & {\footnotesize (\ref{eq:Tmunu_decomp})}\tabularnewline
\midrule 
\textsf{\textbf{\textit{\footnotesize $\pi$}}} & \textsf{\textbf{\textit{\footnotesize $\delta\pi$}}} & {\footnotesize Shear viscosity potential (anisotropic stress)} & {\footnotesize (\ref{eq:potentials})}\tabularnewline
\midrule 
 & $\Theta$ & {\footnotesize Spatial divergence of comoving velocity perturbation} & {\footnotesize (\ref{eq:ThetaX_def})}\tabularnewline
\midrule 
 & $\Xi$ & {\footnotesize Spatial divergence of total energy flow} & {\footnotesize (\ref{eq:XiDef})}\tabularnewline
\midrule 
$c_{\text{s}}^{2}$ &  & {\footnotesize Physical sound speed of dark energy} & {\footnotesize (\ref{eq:cs2})}\tabularnewline
\midrule 
$c_{\text{a}}^{2}$ &  & {\footnotesize Adiabatic sound speed, $\dot{\mathcal{P}}/\dot{\mathcal{E}}$} & {\footnotesize (\ref{eq:EvolvSubst})}\tabularnewline
\midrule 
$C^{2}$ &  & {\footnotesize Coefficient of $\delta\mathcal{E}$ in pressure perturbation
$\delta\mathcal{P}$} & {\footnotesize (\ref{eq:k-ess-closure}),(\ref{eq:ClosureParam})}\tabularnewline
\bottomrule
\end{tabular}\caption{Compilation of notation used in the paper. \label{tab:Notation}}
\end{table}

\section{Covariant decomposition of energy-momentum tensor\label{sec:CovDecomp}}

We decomposed the energy momentum tensor for dark energy according
to
\[
T_{\mu\nu}^{X}=\mathcal{E}u_{\mu}u_{\nu}+\mathcal{P}\perp_{\mu\nu}+2q_{(\mu}u_{\nu)}+\tau_{\mu\nu}\,,
\]
in the scalar frame defined by the vorticity-free velocity field
\begin{equation}
u_{\mu}=-\frac{\nabla_{\mu}\phi}{m}\,.\label{eq:scalframe2}
\end{equation}
We can obtain evolution equations for the fluid by taking the divergence
of the EMT. It is a vector equation, which again can be decomposed
into a time and spatial part for the observer moving with velocity
(\ref{eq:scalframe2}):
\begin{align}
-u^{\nu}\nabla^{\mu}T_{\mu\nu}^{X} & =\frac{\mathrm{d}\mathcal{E}}{\mathrm{d}\tau}+\theta(\mathcal{E}+\mathcal{P})+a^{\mu}q_{\mu}+\nabla_{\mu}q^{\mu}+\sigma_{\mu\nu}\tau^{\mu\nu}\,,\label{eq:Tcons_u}\\
\perp_{\lambda}^{\nu}\nabla^{\mu}T_{\mu\nu}^{X} & =a_{\lambda}\left(\mathcal{E}+\mathcal{P}\right)+\boldsymbol{\overline{\nabla}}_{\lambda}\mathcal{P}+\frac{4}{3}\theta q_{\lambda}+\perp_{\lambda\mu}\frac{\mathrm{d}q^{\mu}}{\mathrm{d}\tau}+\sigma_{\lambda\mu}q^{\mu}+\perp_{\lambda\nu}\nabla_{\mu}\tau^{\mu\nu}\,,\label{eq:Tcons_perp}
\end{align}
We have again assumed that the rotation tensor vanishes as a result
of the definition of our velocity field. When the EMT for DE is conserved
both equations (\ref{eq:Tcons_u}) and (\ref{eq:Tcons_perp}) are
identically zero. They then represent energy conservation (continuity
equation) and momentum conservation (Euler equation) in the frame
comoving with $u^{\mu}$. 

In this discussion, we are interested in analysing scalar-field theories.
Since the scalar-field is isolated, both the energy flow and the viscous-stress
tensor are formed from appropriate derivatives of some scalar-field
potentials, namely
\begin{align}
q_{\lambda} & =\perp_{\lambda}^{\mu}\nabla_{\mu}q=\overline{\boldsymbol{\nabla}}_{\lambda}q\,,\label{eq:q-scalar}\\
\tau_{\mu\nu} & =-\left(\perp_{\mu}^{\alpha}\perp_{\nu}^{\alpha}-\frac{1}{3}\perp_{\mu\nu}\perp^{\alpha\beta}\right)\nabla_{\alpha}\nabla_{\beta}\pi=\label{eq:pi-scalar}\\
 & =-\left(\overline{\boldsymbol{\nabla}}_{\nu}\overline{\boldsymbol{\nabla}}_{\mu}-\frac{1}{3}\perp_{\mu\nu}\triangle\right)\pi+\frac{\mathrm{d}\mathcal{\pi}}{\mathrm{d}\tau}\sigma_{\mu\nu}+u_{\mu}\sigma_{\nu}^{\beta}\overline{\boldsymbol{\nabla}}_{\beta}\pi+\frac{1}{3}\theta u_{\mu}\overline{\boldsymbol{\nabla}}_{\nu}\pi\,,\nonumber 
\end{align}
where $q$ denotes a potential for the energy flow while $\pi$ is
the anisotropic stress, which plays the role of a potential for the
viscous-stress tensor. They are scalar functions of the covariantly
defined variables $\phi$, $m$ (the value of ``velocity'' of the
scalar) and potentially other scalars defined on the spatial hypersurface,
say $\theta$ or $^{(3)}R$, the intrinsic curvature of the three-slice
(see eq.~(\ref{eq:R3_def})). The symbol $\triangle\equiv\perp^{\mu\nu}\overline{\boldsymbol{\nabla}}_{\mu}\overline{\boldsymbol{\nabla}}_{\nu}$
is the Laplacian on the spatial hypersurface. 

Admittedly, such a simplification may not always be quite possible:
the model's viscous-stress tensor could contain some uncontracted
function of the curvature tensor $^{(3)}R_{\mu\nu}$. One may need
to deal with these intrinsically non-scalar terms separately. The
validity of such a decomposition should be confirmed on a model-by-model
basis. 

We should note at this stage that a covariant spatial derivative $\overline{\boldsymbol{\nabla}}_{\mu}$
acting on a function of only $\phi$ vanishes (since\textbf{ $\overline{\boldsymbol{\nabla}}_{\mu}f(\phi)=f_{\phi}\perp_{\mu}^{\nu}\nabla_{\nu}\phi=-mf_{\phi}\perp_{\mu}^{\nu}u_{\nu}=0$},
i.e. since the gradient of $\phi$ defines the time direction). In
many models, such as $f(R)$ gravity or the non-minimally coupled
k-\emph{essence }model discussed in section \ref{sec:BDK}, the anisotropic
stress is only a function of $\phi$ and therefore the viscous stress
tensor reduces to 
\begin{equation}
\tau_{\mu\nu}=\frac{\mathrm{d}\mathcal{\pi}(\phi)}{\mathrm{d}\tau}\sigma_{\mu\nu}\,,\label{eq:tau_phionly}
\end{equation}
and the function $\nicefrac{\mathrm{d}\mathcal{\pi}}{\mathrm{d}\tau}$
plays the role of shear viscosity. Note that depending on the direction
of evolution of $\phi$, this shear viscosity can be of either sign.
This is a hint that we are not dealing with a usual fluid.

Assuming eqs (\ref{eq:q-scalar}) and (\ref{eq:pi-scalar}), we can
re-express the EMT conservation equations (\ref{eq:Tcons_u}) and
(\ref{eq:Tcons_perp}) for the case where the substitution of (\ref{eq:q-scalar})
and (\ref{eq:pi-scalar}) is valid 
\begin{align}
-u^{\nu}\nabla^{\mu}T_{\mu\nu} & =\frac{\mathrm{d}\mathcal{E}}{\mathrm{d}\tau}+\theta(\mathcal{E}+\mathcal{P})+2a^{\mu}\boldsymbol{\overline{\nabla}}_{\mu}q+\triangle q+\frac{\mathrm{d}\pi}{\mathrm{d}\tau}\sigma_{\mu\nu}\sigma^{\mu\nu}-\sigma^{\mu\nu}\boldsymbol{\overline{\nabla}}_{\mu}\boldsymbol{\overline{\nabla}}_{\nu}\pi\,,\label{eq:Tcons_u_scalar}\\
\perp_{\lambda}^{\nu}\nabla^{\mu}T_{\mu\nu} & =a_{\lambda}\left(\mathcal{E}+\mathcal{P}+\frac{\mathrm{d}q}{\mathrm{d}\tau}\right)+\boldsymbol{\overline{\nabla}}_{\lambda}\left(\mathcal{P}+\frac{\mathrm{d}q}{\mathrm{d}\tau}\right)+\theta\boldsymbol{\overline{\nabla}}_{\lambda}q+\perp_{\lambda\nu}\nabla_{\mu}\tau^{\mu\nu}\,.\label{eq:Tcons_perp_scalar}
\end{align}
The gradient of viscous stress can be expressed as 
\begin{align}
\perp_{\lambda\nu}\nabla_{\mu}\tau^{\mu\nu}= & \frac{\mathrm{d}\pi}{\mathrm{d}\tau}R_{\mu\nu}\perp_{\lambda}^{\nu}u^{\mu}+\frac{2}{3}\dot{\pi}\overline{\boldsymbol{\nabla}}_{\lambda}\theta+\frac{\mathrm{d}\pi}{\mathrm{d}\tau}a_{\mu}\sigma_{\lambda}^{\mu}+\sigma_{\lambda}^{\mu}\overline{\boldsymbol{\nabla}}_{\mu}\dot{\pi}-\label{eq:visc_cons}\\
 & -a^{\alpha}\overline{\boldsymbol{\nabla}}_{\lambda}\overline{\boldsymbol{\nabla}}_{\alpha}\pi-\frac{2}{3}\overline{\boldsymbol{\nabla}}_{\mu}\triangle\pi+\frac{1}{3}a_{\lambda}\triangle\pi-\phantom{}^{(3)}R_{\lambda}^{\mu}\overline{\boldsymbol{\nabla}}_{\mu}\pi\,,\nonumber 
\end{align}
with the intrinsic curvature of the spatial hypersurface obtained
through the Gauss-Codazzi equation
\begin{equation}
^{(3)}R_{\mu\nu}=\sigma_{\mu\alpha}\sigma_{\nu}^{\alpha}-\frac{1}{3}\theta\sigma_{\mu\nu}-\frac{2}{9}\theta^{2}\perp_{\mu\nu}+R_{\rho\sigma\alpha\beta}\perp^{\alpha\rho}\perp_{\mu}^{\beta}\perp_{\nu}^{\sigma}\,.\label{eq:R3_def}
\end{equation}
If eq. (\ref{eq:tau_phionly}) is satisfied then eq. (\ref{eq:visc_cons})
simplifies further,
\[
\perp_{\lambda\nu}\nabla_{\mu}\tau^{\mu\nu}=\frac{\mathrm{d}\pi}{\mathrm{d}\tau}\left(R_{\mu\nu}\perp_{\lambda}^{\nu}u^{\mu}+\frac{2}{3}\overline{\boldsymbol{\nabla}}_{\lambda}\theta\right)\,.
\]

The advantage of the fully generally covariant decomposition of the
EMT conservation presented in eqs (\ref{eq:Tcons_u_scalar}) and (\ref{eq:Tcons_perp_scalar})
is that it is exact under the circumstances where the substitution
of (\ref{eq:q-scalar}) and (\ref{eq:pi-scalar}) is valid. As we
demonstrate below, obtaining the linear perturbations equations for
a background flat FLRW metric is very simple. However, starting from
these covariant equations one can also easily derive the perturbation
equations for any other background, for example, a cosmology where
the background shear is non-vanishing (Bianchi) or there is some sort
of (potentially non-isotropic) spatial curvature. In particular, it
is easy to see from eq. (\ref{eq:visc_cons}) that for models where
the anisotropic stress is more complicated than in eq. (\ref{eq:tau_phionly})
there is a non-trivial coupling to background spatial curvature which
will contribute to the evolution of perturbations. %
\footnote{For example, if the action for the scalar contains a term $G_{\mu\nu}\nabla^{\mu}\phi\nabla^{\nu}\phi$
\cite{Sushkov:2009hk}.%
} 

We should note that the equation for the conservation of the EMT for
a scalar field is proportional to the equation of motion for the scalar.
This must be so, since the scalar is a single degree of freedom and
providing a second evolution equation would overconstrain it. This
means that for all scalar-field Lagrangians, eq.~(\ref{eq:Tcons_u_scalar})
is equivalent to the equation of motion while eq.~(\ref{eq:Tcons_perp_scalar})
vanishes identically on all configurations. The momentum conservation
equation can in fact be shown to relate the parts of the pressure
and energy density for the scalar, providing a sort of equation of
state or, more correctly, it is the equilibrium Euler relation for
the scalar field (see e.g. refs \cite{Deffayet:2009wt,Pujolas:2011he}).

\section{Choice of frames and boosts\label{sub:boosts}}

We have performed the decomposition of the EMT in what we have called
the scalar frame, with the result that the fluid variables we've obtained
(energy density, pressure, etc.) have the values which would have
been seen by an observer moving with velocity $u^{\mu}$ defined in
eq.~(\ref{eq:u_def}). A different observer will in general observe
a different energy density, etc. (see also the discussion in ref.~\cite{Ballesteros:2011cm}).

The k-\emph{essence} class of models has EMTs of perfect-fluid form.
The energy-flow $q^{\mu}$ in the scalar frame vanishes and thus the
scalar frame is also the Landau-Lifshitz frame (or rest frame). This
makes the choice of frame for the analysis very simple: the scalar
frame is the natural one. However, in general scalar-field models
(for example, see ref.~\cite{Pujolas:2011he} or section \ref{sec:BDK}),
the energy flow $q_{\mu}$ in the scalar frame is non-vanishing. In
these circumstances, the LL frame is distinct from the scalar frame
and one needs to choose. Let us compare the physics in the two frames.

The two frames are related by a (space-dependent) Lorentz boost. However,
in general there isn't a simple transformation between the two frames.
For the purposes of cosmology, $q_{\mu}$ and $\tau_{\mu\nu}$ in
the background vanish because of the symmetry of the FLRW metric,
so we are able to make a perturbative statement about this transformation. 

The Lorentz boost we require is a transformation from the orthogonal
vector pair $(u^{\mu},q^{\mu})$ to the orthogonal pair $(U^{\mu},Q^{\mu})$
defined by the linear transformation
\[
\left(\begin{array}{c}
U^{\mu}\\
\widehat{Q}^{\mu}
\end{array}\right)=\left(\begin{array}{cc}
\cosh\alpha & \sinh\alpha\\
\sinh\alpha & \cosh\alpha
\end{array}\right)\left(\begin{array}{c}
u^{\mu}\\
\widehat{q}^{\mu}
\end{array}\right)\,,
\]
where $\widehat{q}^{\mu}$ is the unit vector in the direction $q^{\mu}$
and the rapidity $\alpha$ is defined as 
\begin{equation}
\alpha^{2}\equiv\frac{q^{\alpha}q_{\alpha}}{\mathcal{E}+\mathcal{P}}\,,\label{eq:Alpha_def}
\end{equation}
or
\begin{equation}
U_{\mu}\simeq u_{\mu}+\frac{q_{\mu}}{\mathcal{E}+\mathcal{P}}\,,\label{eq:Velocity_Transform_LL}
\end{equation}
to first order in $\alpha$. The frame comoving with velocity $U^{\mu}$
is now the LL (``rest'') frame. In particular, the EMT decomposed
in this new frame to leading order in $\alpha$ can be written 
\begin{eqnarray}
T_{\mu\nu} & \simeq & \left(\mathcal{E}-\frac{2q_{\alpha}q^{\alpha}}{\mathcal{E}+\mathcal{P}}\right)U_{\mu}U_{\nu}+\left(\mathcal{P}-\frac{2q_{\alpha}q^{\alpha}}{3\left(\mathcal{E}+\mathcal{P}\right)}\right)\perp_{\mu\nu}^{U}+\label{eq:EMT_LL}\\
 &  & +\left(\tau_{\mu\nu}-\frac{2\left(q_{\mu}q_{\nu}-\frac{1}{3}q_{\alpha}q^{\alpha}\perp_{\mu\nu}^{U}\right)}{\mathcal{E}+\mathcal{P}}\right)\,.\nonumber 
\end{eqnarray}
with $\perp_{\mu\nu}^{U}\equiv g_{\mu\nu}+U_{\mu}U_{\nu}$. This implies
that up to first order in the rapidity $\alpha$, an observer comoving
with velocity $U^{\mu}$ will observe the same energy density, pressure
and viscous stress as one comoving with $u^{\mu}$: we have
\[
T_{\mu\nu}\simeq\mathcal{E}U_{\mu}U_{\nu}+\mathcal{P}\perp_{\mu\nu}^{U}+\tau_{\mu\nu}\,.
\]
Just as required, the energy flow in this frame vanishes. However,
the decomposition (\ref{eq:EMT_LL}) shows that the observed LL fluid
variables are in fact modified with respect to the scalar frame already
at the second order in $\alpha$. 

It is interesting to note at this point that the singularity of the
transformation to the LL frame when $\mathcal{E}=-\mathcal{P}$ appears
only at the linearised level around a homogeneous and isotropic background.
In fact, the singularity does not appear when one considers the full,
non-linear transformation around FLRW, or the linearised transformation
around some background where the imperfect terms do not vanish. The
full transformation of momentum flux together with the rest of the
fluid and kinematical quantities can be found in ref.~\cite[App. B]{Maartens:1998qw}.
One can see that linearising this full transformation of the momentum
flux under a general velocity change given in ref.~\cite{Maartens:1998qw}
and requiring a zero momentum flux in the new frame, we arrive at
a relation equivalent to equation (\ref{eq:Velocity_Transform_LL}). 

One should note at this point that the above is not a gauge-dependent
statement. Frequently the question of gauge and frame choice are conflated
in cosmology. They are in principle very different. Gauges are a choice
of the form of the metric perturbations which are all equivalent as
a result of diffeomorphism invariance of general relativity. They
do not impact observables.\textbf{ }Under the particular choice of
the Newtonian gauge, which we are using in this paper, the perturbed
equations take the form of (\ref{eq:EnCons_sc}) and (\ref{eq:MOmCons_sc}).\emph{
In this gauge, }the two metric potentials $\Phi$ and $\Psi$ happen
to be coincident with the gauge-invariant Bardeen variables, while
the other metric potentials (usually denoted as $B$ and $E$) vanish.
The same is true of the EMT perturbations: in the Newtonian gauge
they are coincident with the gauge-invariant quantities. This means
that these equations have the same form as the gauge-invariant equations.
All one needs to obtain the gauge-invariant equations is to reinterpret
each of the Newtonian-gauge variables as the corresponding gauge-invariant
one.

The choice of frame, on the other hand, is a statement about observables
as measured by particular observers. As can clearly be seen in eq.
(\ref{eq:EMT_LL}), the quantity that the observer moving with $U^{\mu}$
will call the energy density is a scalar, but it is a different scalar
to the energy density seen by $u^{\mu}$. On the level of linear perturbations,
changing the observer is given by a transformation with a very similar
form to a gauge transformation, but in fact describes a physical change
of observables.

Turning to first-order cosmological perturbation theory, we can find
the spatial velocity divergence for $U^{\mu}$ corresponding to (\ref{eq:ThetaX_def}),
\begin{equation}
\Theta_{\text{LL}}\equiv ik_{i}\delta U^{i}\simeq\Theta_{X}-\frac{k^{2}\delta q}{a^{2}\left(\mathcal{E}+\mathcal{P}\right)}+\frac{\dot{q}}{\mathcal{E}+\mathcal{P}}\Theta_{X}\,,\label{eq:ThetaLL_def}
\end{equation}
and therefore we can rewrite the conservation equations (\ref{eq:EnCons_sc})
and (\ref{eq:MOmCons_sc}) in the Landau-Lifshitz frame,
\begin{flalign}
 & \dot{\delta\mathcal{E}}_{\text{LL}}+3H\left(\delta\mathcal{E}_{\text{LL}}+\delta\mathcal{P}_{\text{LL}}\right)+\left(\mathcal{E}+\mathcal{P}\right)\left(\Theta_{\text{LL}}+3\dot{\Phi}\right)=0\,,\label{eq:EnCons_LL}\\
 & \left(\mathcal{E}+\mathcal{P}\right)\left(\dot{\Theta}_{\text{LL}}+2H\Theta_{\text{LL}}-\frac{k^{2}}{a^{2}}\Psi\right)-\frac{k^{2}}{a^{2}}\delta\mathcal{P}_{\text{LL}}+\Theta_{\text{LL}}\dot{\mathcal{P}}-\frac{2}{3}\frac{k^{4}}{a^{4}}\delta\pi_{\text{LL}}=0\,.\nonumber 
\end{flalign}
where the subscript LL on the fluid variables signifies that these
are the quantities observed in the LL frame and just happen to be
to equal to the scalar-frame equivalents up to first order in $\alpha$.
Unsurprisingly all that has happened is that the terms containing
$q$ have disappeared: the form of these equations must be invariant
under Lorentz boosts.

One could at this stage walk away having decided that one should always
perform cosmological perturbation theory in the Landau-Lifshitz (rest)
frame since it appears computationally simpler as a result of having
eliminated the additional variables $q$. However, this is dangerous
for two reasons:
\begin{itemize}
\item The velocity field $U^{\mu}$ is now no longer vorticity-free. The
scalar-frame velocity field was a (normalised) derivative of a scalar
field and thus the antisymmetric rotation (twist) tensor vanished.
This is no longer true for $U^{\mu}$ and therefore the scalar-field
fluid can carry vector perturbations in this frame---albeit only at
higher orders in perturbations. One may not be particularly worried
about this since at higher orders all the perturbation types mix anyway,
but it does provide additional complication, especially with regard
to choosing a good Cauchy surface. One would also need to add the
rotation tensor terms to eqs (\ref{eq:Tcons_u}) and (\ref{eq:Tcons_perp})
and would not be able to perform the decomposition leading to eq.~(\ref{eq:Tcons_u_scalar})
and (\ref{eq:Tcons_perp_scalar}).
\item Possibly more importantly, eq.~(\ref{eq:Alpha_def}) shows that the
rapidity $\alpha$ required to boost to the LL frame diverges when
the equation of state of the fluid approaches $w=-1$. The physical
origin of this is that the vanishing enthalpy of the vacuum cannot
compensate for the energy flow $q_{\mu}$. This means that as the
fluid approaches the vacuum equation of state, the perturbative relations
between the fluid variables in the LL and any other frame breaks down:
in the LL frame, the energy-density functional form itself contains
corrections (the leading one can be seen in eq.~(\ref{eq:EMT_LL})),
which are order one close to $w=-1$. This also means that order-one
vorticity would be observed in the LL frame.
\end{itemize}
The last point implies that if there is any danger of having an equation
of state close to that of vacuum, one should \emph{not }decompose
the EMT in the LL frame and attempt to evolve these hydrodynamic variables.
On the other hand, any \emph{other }choice of frame is in fact equivalent
up to first order in the difference in velocities and for the purpose
of linear perturbation theory one can directly use results from one
frame and apply them to others. 

One could of course claim that such a setup is not possible: after
all one calculates the hydrodynamical EMT by averaging quantities
in the rest frame of the particles making up the fluid in the first
place. Why should it be possible to produce an EMT which one could
not put back in the rest-frame? As we argue in section \ref{sub:k-ess},
general scalar-field theories do not have EMTs which behave as in
hydrodynamics and indeed have been shown to be able to produce a fluid
vacuum configuration than nonetheless carries momentum in the spatial
direction (e.g.~ref.~\cite{Pujolas:2011he})

It is worth noting here that the practice assigning the meaning of
e.g. the energy density to components of the EMT can also be understood
in the framework of the decomposition (\ref{eq:Tmunu_decomp}). First,
one should realise that it is \emph{impossible} to find a frame where
the definition of energy density is \emph{exactly} $-T_{\phantom{0}0}^{0}$
given an arbitrary metric. The best we can do is to pick the time
direction to be that of the coordinate time of the background, i.e.
\[
u_{\text{com}}^{\mu}=\sqrt{-g^{00}}\delta_{0}^{\mu}\,,
\]
where the square root of the metric is necessary for the appropriate
normalisation of the velocity. In this frame we can then obtain the
following decomposition of the EMT:
\begin{align}
\mathcal{E}_{\text{com}} & =-g^{00}T_{00}=-T_{\phantom{0}0}^{0}+g^{0i}T_{i0}\,,\\
\mathcal{P}_{\text{com}} & =\frac{1}{3}\left(T_{\mu}^{\mu}-\mathcal{E}\right)\,,\nonumber \\
q_{\lambda}^{\text{com}} & =\delta_{\lambda}^{i}\sqrt{-g^{00}}\left(T_{0i}+g_{0i}\mathcal{E}\right)\,,\nonumber \\
\tau_{\mu\nu}^{\text{com}} & =\delta_{\mu}^{i}\delta_{\nu}^{j}\left(T_{ij}-\mathcal{P}g_{ij}-g^{00}\left(\mathcal{E}-\mathcal{P}\right)g_{0i}g_{0j}-g^{00}\left(T_{0i}g_{0j}+T_{0j}g_{0i}\right)\right)\,.\nonumber 
\end{align}
This then matches the components approach, but only provided that
the $(0i)$ metric elements vanish. Even in this case there is an
additional metric factor multiplying $q_{\lambda}$. However, given
scalar perturbation of FLRW in the Newtonian gauge and neglecting
second-order corrections, this decomposition yields the same energy
density, pressure and viscous stress as the decomposition in the scalar
frame.

Another aspect worth mentioning is what happens in the case of an
external EMT. For example, if we consider the scalar field to model
dark energy, dark matter would be an additional external perfect fluid.
The external EMT will contain its own rest frame with velocity, say,
$v^{\mu}$
\[
T_{\mu\nu}^{\text{ext}}=\rho_{\text{ext}}v_{\mu}v_{\nu}+p_{\text{ext}}\perp_{\mu\nu}^{V}\,,\qquad\perp_{\mu\nu}^{V}\equiv g_{\mu\nu}+v_{\mu}v_{\nu}\,.
\]
However, we have to pick just one frame in which to do the decomposition,
and it is convenient to make that the scalar frame. This means that
the fluid variables describing the dark matter are not those observed
in its own rest frame, but appropriately transformed quantities. For
example, the external pressure observed in the frame $u^{\mu}$ would
then be 
\begin{eqnarray}
p_{\text{ext}}^{U} & = & \frac{1}{3}p_{\text{ext}}\left(2+(u\cdot v)^{2}\right)-\frac{1}{3}\rho_{\text{\text{ext}}}\left(1-(u\cdot v)^{2}\right)\,.\label{eq:ExtPres}
\end{eqnarray}
The various properties of the external perfect fluid are defined in
its own rest frame (just as they are for k-\emph{essence}, for example).
Thus in general one needs to keep track of how the external matter
appears in the frame in which we are decomposing and perform the perturbations.
For example, eq.~(\ref{eq:ExtPres})\textbf{ }show that external
dust, which has an equation of state $w_{\text{ext}}=0$ in its own
frame, would have a non-vanishing pressure in the scalar frame.

However, just as in the case of the perturbations of the scalar, if
both the fluids share the FLRW background, the corrections to energy
density, pressure and viscous stress arising from the difference in
the rest-frames appear only at second order, thus for example the
energy density perturbations $\delta\rho_{\text{ext}}$ in the external
rest-frame is going to be the same as in the scalar frame and we can
drop the superscript $U$ used in eq.~(\ref{eq:ExtPres}) to differentiate
the two. At linear order the only difference will be in the energy
flow: since the two fluids do not share the same rest frame, the external
matter will have a non-vanishing velocity from the point of view of
the scalar:
\begin{equation}
\delta q_{\lambda}^{\text{ext}}\equiv T_{\mu\nu}^{\text{ext}}u^{\mu}\perp_{\lambda}^{\nu}\simeq(\rho_{\text{ext}}+p_{\text{ext}})\delta_{\lambda}^{i}\left(\delta v_{i}-\delta u_{i}\right)\,.
\end{equation}
or if we take the divergence,
\[
ik_{i}\delta q_{\text{ext}}^{i}=(\rho_{\text{ext}}+p_{\text{exp}})\left(\Theta_{\text{ext}}-\Theta\right)\,,
\]
where $\Theta_{\text{ext}}=\partial_{i}\delta v^{i}$ is the analogously
defined spatial divergence of the external matter velocity perturbation.

\smallskip{}

To conclude: the choice of frame is in general important when one
is worried about higher orders in cosmological perturbation theory.
If we only concern ourselves with linear perturbations, then any frame
where the observers while unperturbed are comoving with the FLRW background
observers will give the same results. However, for scalar-field models
with second-derivatives in the EMT, the Landau-Lifshitz (rest) frame
is singular when $w=-1$ and perturbation theory loses validity as
the fluid approaches the vacuum-energy equation of state. Thus, somewhat
counter intuitively to expectations, one should use any frame but
the rest frame to study linear perturbations in cosmology.

\bibliographystyle{utcaps}
\bibliography{fluids}

\end{document}